\documentclass[aps,prd,reprint,groupedaddress,superscriptaddress,preprintnumbers]{revtex4-2}

\usepackage{subfigure}
\usepackage{graphicx,natbib}
\usepackage{bm}
\usepackage{color}
\usepackage{amssymb,amsmath}
\usepackage[margin=0.6in]{geometry}
\usepackage{multirow}
\usepackage{mathrsfs}
\usepackage{hyperref}

\newcommand{\meanBB}{\langle\mbox{\boldmath $B$}\rangle{}}{}
\newcommand{\alphaem}{\ensuremath{\alpha_{\rm em}}}

\newcommand{\meanmufive}{\bra{\mu_5}}
\newcommand{\meanEMF}{\bra{\mbox{\boldmath{${\cal{E}}$}}}{}}{}

\newcommand{\BB}{\bm{B}}

\newcommand{\UU}{\bm{U}}

\newcommand{\EE}{\bm{E}}

\newcommand{\SSSS}{\mbox{\boldmath ${\sf S}$} {}}

\newcommand{\meanrho}{\langle{\rho}\rangle}
\def\cs{c_{\rm s}}
\newcommand{\nab}{\bm{\nabla}}
\def\Rm{{\rm Re}_{_\mathrm{M}}}
\newcommand{\bra}[1]{\langle #1\rangle}
\def\cs{c_{\rm s}}

\def\kB{k_{\rm B}}

\newcommand{\jcap}{J. Cosmol. Astropart. Phys.}

\begin{document}

\title{Efficiency of dynamos from an autonomous generation of chiral asymmetry}
\preprint{NORDITA-2024-008}

\author{Jennifer~Schober}
\email{jennifer.schober@epfl.ch}
\affiliation{Institute of Physics, Laboratory of Astrophysics, \'Ecole Polytechnique F\'ed\'erale de Lausanne (EPFL), 1290 Sauverny, Switzerland}

\author{Igor~Rogachevskii}
\affiliation{Department of Mechanical Engineering, Ben-Gurion University of the Negev, P.O. Box 653, Beer-Sheva 84105, Israel}
\affiliation{Nordita, KTH Royal Institute of Technology and Stockholm University, 10691 Stockholm, Sweden}

\author{Axel~Brandenburg}
\affiliation{Nordita, KTH Royal Institute of Technology and Stockholm University, 10691 Stockholm, Sweden}
\affiliation{The Oskar Klein Centre, Department of Astronomy, Stockholm University, AlbaNova, SE-10691 Stockholm, Sweden}
\affiliation{School of Natural Sciences and Medicine, Ilia State University, 0194 Tbilisi, Georgia}
\affiliation{McWilliams Center for Cosmology and Department of Physics, Carnegie Mellon University, Pittsburgh, Pennsylvania 15213, USA}

\date{\today}

\begin{abstract}
At high energies, the dynamics of a plasma 
with charged fermions can be described 
in terms of chiral magnetohydrodynamics.
Using direct numerical simulations, we demonstrate
that chiral magnetic waves (CMWs)
can produce a chiral asymmetry $\mu_5 = \mu_\mathrm{L} - \mu_\mathrm{R}$ from a spatially fluctuating (inhomogeneous) chemical potential
$\mu = \mu_\mathrm{L} + \mu_\mathrm{R}$, where
$\mu_\mathrm{L}$ and $\mu_\mathrm{R}$ are the
chemical potentials of left- and right-handed electrically charged fermions, respectively.
If the frequency of the CMW is less than or
comparable to the characteristic growth rate of the 
chiral dynamo instability, the magnetic field can be 
amplified on small spatial scales.
The growth rate of this small-scale chiral dynamo instability is determined by the spatial maximum value of $\mu_5$ fluctuations.
Therefore, the magnetic field amplification occurs during periods when $\mu_5$ reaches temporal maxima during the CMW.
If the small-scale chiral dynamo instability leads to a magnetic field strength that exceeds a critical value, which depends on the resistivity and the initial value of $\mu$, 
magnetically dominated turbulence is produced.
Turbulence gives rise to a large-scale dynamo instability, which we find to be caused by the magnetic alpha effect. 
Our results have consequences for the dynamics of certain high-energy
plasmas, such as the early Universe. 
\end{abstract}

\maketitle

\section{Introduction}

In the Standard Model of particle physics,
the chirality of high-energy fermions
can lead to macroscopic quantum effects,
which are a result of the chiral anomaly.
A prominent example is the chiral magnetic effect (CME)
\citep{Vilenkin:80a}, which is relevant at high energies and can lead to a 
magnetic-field-aligned electric current, if there is an asymmetry
between the number density of left- and right-handed electrically charged fermions.
The emergence of the CME and other novel quantum phenomena 
in nonequilibrium relativistic quantum matter can be derived from first principles
\citep{RW85,Tsokos:85,AlekseevEtAl1998,Frohlich:2000en, Frohlich:2002fg,Fukushima:08,Son:2009tf, KH14}.
However, to improve the usability of the models, lots of effort has been put into the development
of a quantum kinetic theory for massless fermions often referred to as 
chiral kinetic theory
\citep{SonYamamoto2012,StephanovYin2012,Gorbar:2016klv, HidakaEtAl2022}.
The additional electric current caused by the CME can also be incorporated into an effective description of a relativistic plasma. 
Such models have become known as 
chiral (or anomalous) magnetohydrodynamics (MHD)
\citep{GI13,BFR15,YA16,RogachevskiiEtAl2017,HattoriEtAl2019}.
This paper is based on chiral MHD 
as its theoretical framework.

Chiral phenomena occur in plasmas with fermions that
are effectively massless.
In the context of astrophysics and cosmology 
(see \citep{KamadaEtAl2023} for a recent review),
this limits the applications to high-energy plasma in 
which the temperature is above $10$ MeV \citep{BFR12}. 
A prime example is the hot and dense plasma that fills the early Universe. 
It was first suggested in Ref.~\citep{JS97} that the CME can lead to an instability in the primordial magnetic field, which is now known as the chiral plasma instability \cite{KamadaEtAl2023} or the small-scale
chiral dynamo instability
\footnote{This instability is referred to differently in the literature of different fields.
In high-energy physics, it is often called the ``chiral plasma instability'', but also the names ``$v_{\mu}$-'' or ``$v_{5}$-''dynamo exist. In this paper, 
we only use the notion ``small-scale chiral dynamo instability''.}.
If the dynamo is excited, strong helical magnetic fields can be generated \citep{TashiroEtAl2012},
which can drive magnetically dominated 
turbulence that gives rise to mean-field dynamos \citep{RogachevskiiEtAl2017,SchoberEtAl2017}.
These primordial magnetic fields can potentially explain the baryon asymmetry of the Universe
\citep{FujitaKamada2016,KamadaLong2016}, produce relic gravitational waves
\citep{RoperPolEtAl2020,BrandenburgEtAl2024}, and affect the properties 
of the global 21 cm signal \citep{Kunze2019} and
dwarf galaxies \citep{SanatiEtAl2024}.

A second domain within astrophysics and cosmology where 
the chiral anomaly becomes relevant is core-collapse 
supernovae. Here, a chiral imbalance 
is generated through the emission of neutrinos which are, in the 
Standard Model of particle physics, only left-handed. 
Chiral effects have been included in modeling
the magnetic field evolution in core-collapse supernovae \citep{MasadaEtAl2018,MatsumotoEtAl2022}, 
and were suggested to play a role in the generation of 
magnetars \cite{Ohnishi:2014uea,DS15,Dvornikov:2015lea, SiglLeite2016} and the occurrence of pulsar kicks \citep{CharbonneauEtAl2010, KaminskiEtAl2014}.
These ideas have recently been extended by possible implications of the chiral anomaly in magnetospheres of pulsars \citep{GS22}, where the produced chiral asymmetry can be substantial.
It can trigger the small-scale chiral dynamo which, in turn, can produce circularly polarized electromagnetic radiation in a wide range of frequencies, spanning from radio to near-infrared.
This can affect some features of fast radio bursts.

Beyond the extreme environments in the Universe, chiral effects can be 
studied more directly in heavy ion colliders \citep{K16}. 
However, the existence of the CME has not yet been confirmed in experiments conducted at the Large Hadron Collider \citep{ALICE2013}
or the Relativistic Heavy Ion Collider \citep{STAR2021}.
At low energies, chiral effects can emerge in new materials that include massless 
quasiparticles \citep{MS15,AMV18,GMSS2021}.
The detection of the pseudorelativistic analogues
of CME are realized by the low-energy electron
quasiparticles in Dirac and Weyl materials 
\citep{VV14,LiEtAl2016,LinEtAl2024}, and it
opens up the possibilities of novel technological developments 
\citep[e.g., in the field of quantum computing][]{KharzeevLi2019}.

Chiral MHD differs from classical MHD by an additional 
term in the induction equation, which describes the evolution of the magnetic field. 
This term stems directly from the additional contribution to the electric current from the CME and is proportional to the 
chiral chemical potential $\mu_5 \equiv \mu_\mathrm{L} - \mu_\mathrm{R}$,
where $\mu_\mathrm{L}$ and $\mu_\mathrm{R}$ are the chemical potentials of left- and right-handed fermions, respectively. 
This additional term leads to an instability in the magnetic field on small spatial scales \citep{JS97}, 
the (small-scale) chiral dynamo instability, 
if $\mu_5$ is nonzero.
The amplification of magnetic energy 
in the nonlinear stage of the chiral dynamo instability
can cause the production of magnetically dominated 
turbulence, making exact analytical treatment unfeasible. 
Nevertheless, mean-field theory allows for exploring 
the effects of turbulence in chiral MHD.
In particular, the occurrence of a new mean-field dynamo, i.e., 
the $\alpha_\mu$ dynamo, was predicted in Ref.~\citep{RogachevskiiEtAl2017}.
With direct numerical simulations (DNS), it has been shown that, 
in the nonlinear evolutionary stage, a mean-field dynamo instability can occur
\citep{SchoberEtAl2017}.
In recent studies \citep{SchoberEtAl2022a,SchoberEtAl2022b}, it has
been demonstrated that the chiral dynamo instability even occurs in a
plasma with an initial spatial fluctuating chiral chemical potential
with zero mean.
A necessary condition for a chiral dynamo instability is that the effective 
correlation length of chiral chemical potential fluctuations is larger than the corresponding instability length scale,
which is given by the inverse of the spatial maximum value of $\mu_5$.

The aforementioned studies have explored the role of the CME in the 
evolution of magnetic fields.
However, the CME is not the only macroscopic quantum effect that results from the chiral anomaly.
Another prominent example is the chiral separation effect (CSE) 
\citep{SonZhitnitsky2004,MZ05}.
The CSE is a complementary transport phenomenon to the CME in which a nonzero chemical potential $\mu = \mu_\mathrm{L} + \mu_\mathrm{R}$ 
generates an axial current along an external magnetic field.
A consequence of the CSE is the possibility of exciting new collective modes, most notably the chiral magnetic wave (CMW) \citep{KY11}.
These waves imply periodic conversion between $\mu_5$ 
and $\mu$, in the presence of a small background magnetic field 
and nonvanishing gradients of $\mu_5$ and $\mu$.
Chiral magnetic waves in chiral plasma have been studied 
in a number of publications 
\citep{ZhouXu2018,RGS19,IkedaKharzeevShi2023,AhnEtAl2024}.
Simulating the CMW in a Cartesian domain, it has been recently shown \citep{SchoberEtAl2024}
that the chiral dynamo instability and even mean-field 
dynamos can occur for vanishing initial chiral asymmetry 
if initial spatial fluctuations of the chemical potential are inhomogeneous $({\bm \nabla} \mu \not=0)$.
In this study, we explore the parameter space of CMWs (for which the
chemical potential is nonuniform) and identify the conditions under which
the chiral dynamo instability and mean-field dynamos can be 
excited for plasmas with vanishing initial chiral asymmetry.

The outline of this paper is as follows.
In Sec.~\ref{sec_background} we present the system of equations 
that describe plasma with relativistic fermions including the 
CME and CSE, and discuss the initial conditions that we consider. 
In Sec.~\ref{sec_phenomenology} the evolution of the system is modeled 
phenomenologically and we make some predictions for different scenarios. 
The system of equations is solved numerically in
Sec.~\ref{sec_DNS}, where we compare our predictions
with the numerical results. 
Finally, the results are discussed in Sec.~\ref{sec_discussion} and conclusions are drawn in Sec.~\ref{sec_conclusion}.

\section{System of equations}
\label{sec_background}

\subsection{Chiral MHD equations with CSE}
\label{ChiralMHD}

In this paper, we study effects of relativistic fermions 
applying an effective fluid description for plasma motions. 
As in our previous study \cite{SchoberEtAl2024}, we consider the following set of equations which includes both the 
CME and the CSE (see Appendix~\ref{sec_justification}):
\begin{eqnarray}
  \frac{\partial \BB}{\partial t} &=& \nab   \times   \left[{\UU}  \times   {\BB}
  + \eta \, \left(\mu_5 {\BB} - \nab\times{\BB}\right) \right] ,
\label{ind-DNS_CSE}\\
  \rho{D \UU \over D t}&=& (\nab   \times   {\BB})  \times   \BB
  -\nab  p + \nab  {\bm \cdot} (2\nu \rho \SSSS) ,
\label{UU-DNS_CSE}\\
  \frac{D \rho}{D t} &=& - \rho \, \nab  \cdot \UU ,
\label{rho-DNS_CSE}\\
  \frac{D \mu}{D t}  &=& -\mu \, \nab  \cdot \UU 
  - \mathcal{D}_\mu \, \nabla^4 \mu - C_\mu (\BB {\bm \cdot} \nab)  \mu_5,
  \label{mu-DNS_CSE}  \\
  \frac{D \mu_5}{D t} &=& 
  -\mu_5 \, \nab  \cdot \UU
- \mathcal{D}_5 \, \nabla^4 \mu_5
 - C_5 ({\BB} {\bm \cdot} \nab) \mu \nonumber \\
  &&    + \lambda \, \eta \, \left[{\BB} {\bm \cdot} (\nab   \times   {\BB})
  - \mu_5 {\BB}^2 \right] .
\label{mu5-DNS_CSE} 
\end{eqnarray}
Here, $\BB$ and $\UU$ are the magnetic field and the velocity field, respectively, 
$\eta$ is the microscopic magnetic diffusivity,
$p$ is the pressure, $\nu$ is
the viscosity,
$\rho$ is the mass density, $\SSSS$ is the trace-free 
strain tensor with components
${\sf S}_{ij}=(\partial_j U_i+\partial_i U_j)/2 -
\delta_{ij} ({\bm \nabla}{\bm \cdot} \UU)/3$.
In Eq.~(\ref{mu5-DNS_CSE}), $\lambda=3 \hbar c (8 \alphaem / \kB T)^2$ is the chiral feedback parameter,
where $\hbar$ is the reduced Planck constant, $c$ is the speed of light, 
$\alphaem \approx 1/137$ is the fine structure constant, $\kB$ is the Boltzmann constant, and $T$ is the temperature.
To close the system of equations, we use an isothermal equation of state 
$p = \rho c_{\rm s}^2$, where $c_{\rm s}$ is the sound speed.
For numerical stability, the evolution equations for 
$\mu_5$ and $\mu$ also include (hyper)diffusion terms 
with the diffusion coefficients 
$\mathcal{D}_5$ and $\mathcal{D}_\mu$ \citep{SchoberEtAl2022b}.
The coupling between $\mu_5$ and $\mu$, the strength 
of which is determined by the coupling constants $C_5$ and $C_\mu$, 
leads to CMWs \citep{KY11}.
When considering the coupled linearized equations~(\ref{mu-DNS_CSE})
and (\ref{mu5-DNS_CSE}), the frequency of CMWs is found to be
\begin{eqnarray}
  \omega_{\rm CMW} =  \pm \left[C_5 \, C_\mu 
   ({\bm k} \cdot {\bm B}_\mathrm{ex})^2 
       - {1 \over 4} \left(\lambda \eta \,B_\mathrm{ex}^2
                    \right)^2 \right]^{1/2} ,
\label{eq_omegaCMW}
\end{eqnarray}
where ${\bm B}_\mathrm{ex}$ is the external magnetic field
and ${\bm k}$ is the wave vector.
As long as the magnetic fluctuations are
smaller than ${\bm B}_\mathrm{ex}$, 
the characteristic timescale 
of these waves is half of the period 
\begin{eqnarray}
  P_\mathrm{CMW}= \frac{2\pi}{\omega_\mathrm{CMW}},
\label{eq_PCMW}
\end{eqnarray}
since this is the timescale on which the sign of $\mu_5$ changes.
The damping rate of the CMW is
\begin{eqnarray}
  \gamma_{\rm CMW} = - {1 \over 2} \lambda \eta \,B_\mathrm{ex}^2 .
\label{damp_CMW}
\end{eqnarray}

\subsection{Initial conditions}
\label{sec_IC}

We consider initial conditions where $\mu_5(t_0) = 0$ and $\mu(t_0)$
are spatially random fields consisting of Gaussian noise with a power law spectrum, $E_\mu(k,t_0) \propto \left(k/k_1\right)^{s}$, where $k_1$ is the minimum 
wave number in the system.  The initial magnetic field is weak and in the form of Gaussian noise.
Additionally, we consider an external very weak uniform magnetic field
with $\BB_\mathrm{ex}=(B_\mathrm{ex}, 0, 0)$ to support CMWs, which effectively produce the chiral asymmetry, i.e., a difference in the left-
and right-handed chemical potentials.
The initial velocity field vanishes.

\section{Phenomenology}
\label{sec_phenomenology}

In this section, we discuss the evolution of a plasma with 
inhomogeneous chemical potential phenomenologically. 
We describe the linear phase of the production of $\mu_5$ from 
the chiral separation effect in Sec.~\ref{sec_production}. 
Since we consider a system with an imposed magnetic field,
the produced inhomogeneous $\mu_5$ necessarily leads to an effect 
that we call ``chiral tangling'', as we describe in 
Sec.~\ref{sec_chiral_tangling}. 
If $\mu_5$ exceeds a critical value, the small-scale dynamo 
instability is excited and the magnetic field grows 
exponentially, as discussed in Sec.~\ref{sec_CDI}.
The magnetic field drives turbulence which, if the Reynolds number becomes larger than unity, can give rise to a mean-field dynamo instability,
amplifying the field on large spatial scales.  
The physics of the
mean-field dynamo is described in Sec.~\ref{sec_LSD}.

\subsection{Production of $\mu_5$}
\label{sec_production}

In the initial phase, a chiral asymmetry is generated 
via the term involving $C_5$ in Eq.~(\ref{mu5-DNS_CSE}).
For times less than the period of a CMW, i.e., $t \ll  2\pi \omega_\mathrm{CMW}^{-1}$,
the evolution of $\mu_5$ can be approximated as
\begin{eqnarray}
  \mu_5(t) \approx - C_5 ({\BB_\mathrm{ex}} {\bm \cdot} \nab) \mu \, t.
\end{eqnarray}
Assuming an initial condition 
where $\mu(t_0)$ has a characteristic 
wave number $k_{\mu,\mathrm{eff}}$, we find 
\begin{eqnarray}
  |\mu_5(t)| \approx - C_5 B_{\mathrm{ex}} k_{\mu,\mathrm{eff}} \mu(t_0) \, t.
\label{eq_mu5_init_gen}
\end{eqnarray}
Note that, even for $t \ll 2\pi \omega_\mathrm{CMW}^{-1}$, 
$\mu$ can be a function of time due to the dissipation term in Eq.~(\ref{mu-DNS_CSE}).

Given that the chemical potential has a 
spectrum $E_\mu \propto k^s$, we can write for its $k$-dependent value
\begin{eqnarray}
    \mu^2(k) \approx E_\mu(k) k  \propto  k^{1+s} .
\end{eqnarray}
Inserting $\mu(k)\approx  k^{(1+s)/2}$
in Eq.~(\ref{eq_mu5_init_gen}), yields
\begin{eqnarray}
    \mu_5(k,t) &\propto&  C_5 B_\mathrm{ex}   
    k^{(3+s)/2} ~ t
\end{eqnarray}
The spectrum of $\mu_5$ is then
\begin{eqnarray}
   E_5(k) = \frac{\mu_5(k)^2}{k} \propto 
   C_5^2 B_\mathrm{ex}^2 k^{2+s} ~t^2.
\label{eq_mu5spec}
\end{eqnarray}

\subsection{Chiral tangling}
\label{sec_chiral_tangling}

Nonuniform fluctuations of the chemical potential $\mu$ produce nonuniform 
fluctuations of the chiral chemical potential $\mu_5$ due
to the term $-C_5 ({\BB} {\bm \cdot} \nab) \mu$ in Eq.~(\ref{mu5-DNS_CSE}).
This can lead to a linear in time growth of 
magnetic fluctuations, which is analogous to tangling of 
an external magnetic field by velocity fluctuations.
The relevant term in the induction equation is 
\begin{eqnarray}
    \frac{\partial \BB}{\partial t} = \eta {\bm \nabla} \mu_5 \times \boldsymbol{B}_\mathrm{ex}.   
\end{eqnarray}
We call this effect ``chiral tangling'' and expect it to be only
relevant in early phases, or in cases where the generation of $\mu_5$
is not efficient enough to lead to a chiral dynamo instability.

\subsection{Small-scale chiral dynamo instability}
\label{sec_CDI}

If $\mu_5$ exceeds a critical value, 
a chiral dynamo instability amplifies the magnetic
field exponentially with the maximum growth rate
\begin{eqnarray}
  \gamma_5= \frac{\eta \mu_{5,\mathrm{max}}^2}{4},
\label{eq_gamma5}
\end{eqnarray}
where $\mu_{5,\mathrm{max}}$ is the spatial maximum of $\mu_5$
\citep{SchoberEtAl2022a,SchoberEtAl2022b}. 
The expression given in Eq.~(\ref{eq_gamma5}) is the maximum possible growth rate and is only reached if the instability wave number
\begin{eqnarray}
  k_5= \frac{\mu_{5,\mathrm{max}}}{2},
\label{eq_k5}
\end{eqnarray}
is much larger than the effective correlation 
wave number $k_{\mu_5,\mathrm{eff}}$
of $\mu_5$ \cite{SchoberEtAl2022a}, where
\begin{eqnarray}
  k_{\mu_5,\mathrm{eff}}^{-1}(t) = \frac{\int E_5(k) k^{-1}~\mathrm{d}k}{\int E_5(k)~\mathrm{d}k}.
\label{eq_kmu5eff}
\end{eqnarray}

Note that during this phase, $\mu_5$ continues to grow
similar to Eq.~(\ref{eq_mu5_init_gen}), but 
the external constant field $B_{\mathrm{ex}}$ 
is being replaced by $B_{\mathrm{ex}}+B_\mathrm{rms}(t)$, 
once $B_\mathrm{rms}(t) \gtrsim B_{\mathrm{ex}}$.
Therefore, the produced chiral chemical 
potential $\mu_5$ depends on the magnetic fluctuations $B_\mathrm{rms}$ and the chiral dynamo instability becomes nonlinear.

Whether a large enough chiral asymmetry can be produced to trigger a chiral dynamo instability depends on the initial $\mu$ as well as on the characteristic 
parameters of the system. 
The first necessary condition for a dynamo is that the maximum value of $\mu$, $\mu_\mathrm{max}$, needs to be 
much larger than 
its effective correlation length, $k_{\mu,\mathrm{eff}}$.
Only then, a large enough $\mu_{5,\mathrm{max}}$ can be produced such that the dynamo instability scale, $\mu_\mathrm{max}/2$, 
exceeds $k_{\mu_5,\mathrm{eff}}$.
The second necessary condition for the dynamo instability in CMWs is that the chiral 
dynamo needs to operate on a timescale that is less than half of the period of the CMW, $P_\mathrm{CMW}/2 = \pi/\omega_\mathrm{CMW}$.
In other words, the (minimum possible) dynamo timescale
\begin{eqnarray}
    t_{\mathrm{D}} = \frac{4}{\eta \mu^2_\mathrm{max}(t_0)}.
\label{eq_tD}
\end{eqnarray}
needs to be shorter than $P_\mathrm{CMW}/2$.
In Eq.~(\ref{eq_tD}) we assume that, 
at times when $\mu_\mathrm{5,max}$ reaches its maxima,
its value corresponds to $\mu_\mathrm{max}(t_0)$
[which implies that the dissipation of $\mu_5$ and $\mu$ is insignificant]
and therefore the maximum possible growth rate $\gamma_5$ is determined by $\mu_\mathrm{max}(t_0)$.

\subsection{Maximum possible magnetic field strength generated by the chiral dynamo}

\begin{figure}[t!]
\centering
\includegraphics[width=0.5\textwidth]{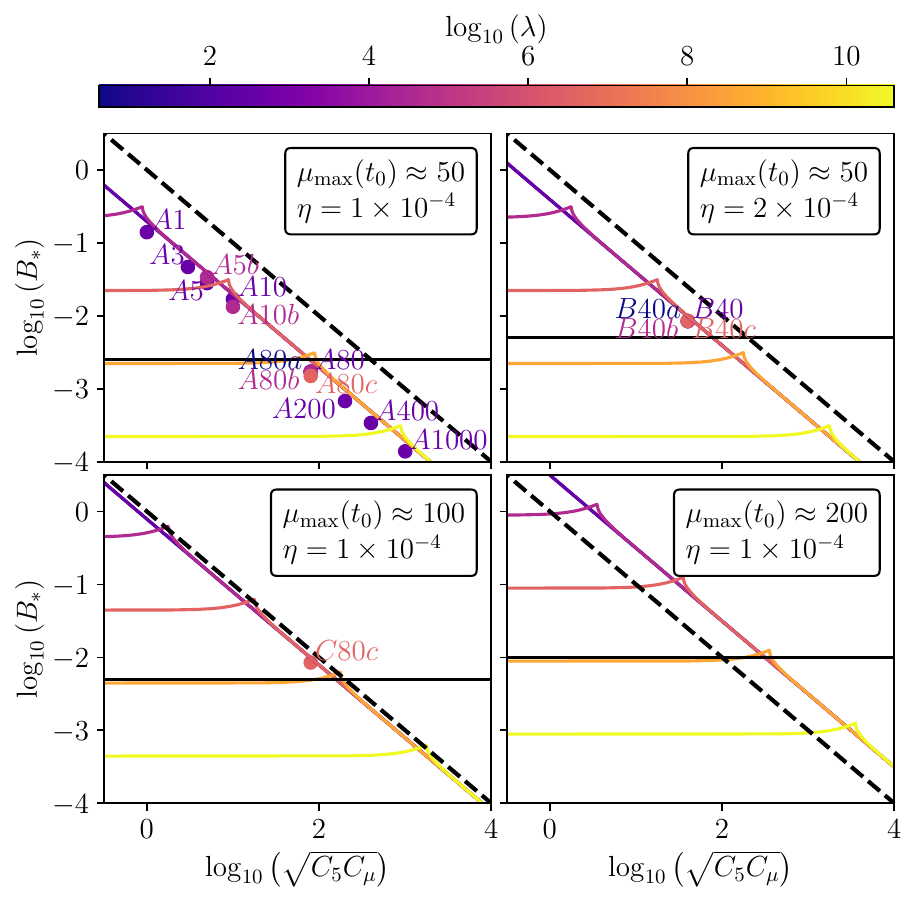}
  \caption{Dependence of the maximum magnetic field strength
  that can result from a chiral magnetic wave, $B_*$
  as given by Eq.~(\ref{eq_BCMW}),
  on the coupling constants $C_5$ and $C_\mu$.
  The different panels indicate different combinations of the initial 
  value of the maximum chemical potential, $\mu_\mathrm{max} (t_0)$, 
  and the magnetic resistivity, $\eta$.
  Colors indicate the value of the feedback parameter $\lambda$.
  The horizontal black lines indicate the approximate threshold 
  for the production of turbulence, as estimated in Eq.~(\ref{eq_Bcrit}),
  and the dashed black line shows the approximate minimum 
  value above which the velocity of the CMW becomes supersonic; see Appendix~\ref{sec_DNScriteria}.
  Overplotted are the 
  values of $B_*$, which are obtained when inserting the parameters 
  for all the runs presented in this
  paper (see Table~\ref{tab_DNSoverview}) into Eq.~(\ref{eq_BCMW}).}
  \label{fig_Bsat_C5Cmu}
\end{figure}

Within one period of the wave, $2\pi/\omega_\mathrm{CMW}$, the sign of 
the produced $\mu_5$ oscillates between
positive and negative.
Therefore a chiral dynamo instability can amplify the magnetic field significantly as long as 
the timescale on which $\mu_5$ changes sign, 
\begin{eqnarray}
    t_{\mathrm{CMW, nl}} &\approx& \frac{\pi}{\left[C_5 \, C_\mu (k_{\mu,\mathrm{eff}} B_\mathrm{rms})^2 
     - {1 \over 4} (\lambda \eta B_\mathrm{rms}^2)^2  
     \right]^{1/2} }   ,\;\;
\label{eq_tCMW}
\end{eqnarray}
is longer than the dynamo timescale in Eq.~(\ref{eq_tD}).
In Eq.~(\ref{eq_tCMW}), we assume that the system is at a stage where magnetic fluctuations are larger than the imposed field. 
With increasing $B_\mathrm{rms}$, $t_{\mathrm{CMW, nl}}$ decreases and 
eventually becomes comparable with $t_{\mathrm{D}}$.
This allows estimating the maximum strength of a magnetic field 
produced by CMWs.
Comparing Eqs.~(\ref{eq_tCMW}) and (\ref{eq_tD}) yields a maximum magnetic field
strength of 
\begin{eqnarray}
|B_*| &=& 
  \frac{\sqrt{2}}{\eta \lambda}  \sqrt{C_5 \, C_\mu} k_{\mu,\mathrm{eff}} \biggl[\pm   
   \left(1 - \xi^2\right)^{1/2} 
   + 1 \biggr]^{1/2}
\label{eq_BCMW}
\end{eqnarray}
with 
\begin{eqnarray}
  \xi \equiv \frac{\pi  \lambda \eta^2 \mu^2_\mathrm{max}(t_0) }{4 C_5 \, C_\mu 
  k_{\mu,\mathrm{eff}}^2}.
\end{eqnarray} 
This expression for $B_*$ is based on the assumptions that 
(i) a complete conversion of $\mu$ to $\mu_5$ is possible and 
(ii) that there is no turbulence in the system.
It is worth noting that Eq.~(\ref{eq_BCMW}) approaches
\begin{eqnarray}
   |B^-_*|  &=& \frac{\pi \eta \mu^2_\mathrm{max}(t_0)}{ 4 \sqrt{C_5 \, C_\mu} k_{\mu,\mathrm{eff}}},
\label{eq_BCMW_highfreq}
\end{eqnarray}
or 
\begin{eqnarray}
   |B^+_*|  &=& \frac{2 \sqrt{C_5 \, C_\mu}k_{\mu,\mathrm{eff}}}{\eta \lambda},
\label{eq_BCMW_highfreq2}
\end{eqnarray}
if $\xi \ll 1$. 
The physically relevant value is 
$|B_*|=\mathrm{min}(|B^-_*|,|B^+_*|)$,
because as soon as the magnetic field strength reaches the lower branch of the solutions, 
the sign of $\mu_5$ changes on a timescale that is shorter than $t_\mathrm{D}$.
In the limit of $\xi \gg 1$ Eq.~(\ref{eq_BCMW}) becomes
\begin{eqnarray}
  |B_*| = \frac{\sqrt{\pi}}{2\sqrt{\lambda}}
          ~\mu_\mathrm{max}(t_0) .
\label{eq_highchi}
\end{eqnarray}
However, in this limit the damping of the CMW can be 
significant, see Eq.~(\ref{damp_CMW}), and $\mu_{5, \mathrm{max}}$ never reaches the maximum possible value of $\mu_{5, \mathrm{max}}= \mu_{\mathrm{max}}(t_0)$.
Therefore, the expression in Eq.~(\ref{eq_highchi}) can be considered as an upper limit. 
The expression given by Eq.~(\ref{eq_BCMW}) is plotted
for different parameters in Fig.~\ref{fig_Bsat_C5Cmu}.

\begin{table*}
\small
\centering
\caption{Summary of the simulations. 
The three reference runs are in bold. 
Runs $A3$ and $A10b$ have been presented also in \citet{SchoberEtAl2024}, where they were named $R1$ and $R2$, respectively. 
Run $R$--$2$ is a comparison run without the CSE, which has been presented in \citet{SchoberEtAl2022b}.
 }
\begin{tabular}{ l | lllll | lll | lll  | llll }
\hline
~     &  \multicolumn{5}{l|}{Parameters}  &  \multicolumn{3}{l|}{Initial conditions} &  \multicolumn{3}{l|}{Phenomenology} &   \multicolumn{4}{l}{DNS results (maximum values)}  \\
Run   &  Res.   & $B_\mathrm{ex}$   &  $ \eta$ &  $\lambda$  &  
$C_5 =C_\mu$    & $\mu_{\mathrm{max}}$            &  $E_\mu$  & $E_5$  
& $\frac{P_\mathrm{CMW}/2}{t_\mathrm{D}}$ & $B_*$  & $B_\mathrm{crit}$  
& $\mu_{\mathrm{5,max}}$ & $B_\mathrm{rms}$  &  $\Rm$  &  $\mathrm{Lu}$   \\
\hline
$A1$        &  $672^3$     &  $1\times 10^{-4}$     &  $1\times 10^{-4}$     &  $4 \times 10^{2}$    &  $1$                  &  $45$       &  $\propto k^{-4}$  &  0                   &  $1400.99$         &  $0.14$        &  $0.0022$       &  $59$      &  $0.041$      &  $58$                 &  $200$    \\
$A3$        &  $1024^3$  &  $1\times 10^{-4}$     &  $1\times 10^{-4}$     &  $4 \times 10^{2}$    &  $3$                  &  $45$       &  $\propto k^{-4}$  &  0                   &  $467.63$          &  $0.047$       &  $0.0023$       &  $68$      &  $0.12$       &  $77$                 &  $440$  \\
$\boldsymbol{A5}$        &  $\boldsymbol{1024^3}$  &  $\boldsymbol{1\times 10^{-4}}$     &  $\boldsymbol{1\times 10^{-4}}$     &  $\boldsymbol{4 \times 10^{2}}$    &  $\boldsymbol{5}$                  &  $\boldsymbol{45}$       &  $\boldsymbol{\propto k^{-4}}$  &  $\boldsymbol{0}$                   &  $\boldsymbol{280.60}$          &  $\boldsymbol{0.028}$       &  $\boldsymbol{0.0023}$       &  $\boldsymbol{63}$      &  $\boldsymbol{0.15}$       &  $\boldsymbol{110}$  &  $\boldsymbol{610}$  \\
$A5b$       &  $720^3$     &  $1\times 10^{-4}$     &  $1\times 10^{-4}$     &  $4 \times 10^{4}$    &  $5$                  &  $48$       &  $\propto k^{-4}$  &  0                   &  $338.65$          &  $0.034$       &  $0.0024$       &  $50$      &  $0.011$      &  $8.4$                &  $40$                 \\
$A10$       &  $720^3$     &  $1\times 10^{-4}$     &  $1\times 10^{-4}$     &  $4 \times 10^{2}$    &  $10$                 &  $48$       &  $\propto k^{-4}$  &  0                   &  $169.32$          &  $0.017$       &  $0.0024$       &  $42$      &  $0.0077$     &  $0.89$               &  $9.3$                \\
$A10b$      &  $672^3$     &  $1\times 10^{-4}$     &  $1\times 10^{-4}$     &  $4 \times 10^{4}$    &  $10$                 &  $44$       &  $\propto k^{-4}$  &  0                   &  $133.03$          &  $0.013$       &  $0.0022$       &  $38$      &  $0.024$      &  $36$                 &  $98$                 \\
$A80a$      &  $448^3$     &  $1\times 10^{-4}$     &  $1\times 10^{-4}$     &  $4$                  &  $80$                 &  $44$       &  $\propto k^{-4}$  &  0                   &  $17.17$           &  $0.0017$      &  $0.0022$       &  $66$      &  $0.00043$  &  $0.054$              &  $0.45$               \\
$\boldsymbol{A80}$       &  $\boldsymbol{448^3}$     &  $\boldsymbol{1\times 10^{-4}}$     &  $\boldsymbol{1\times 10^{-4}}$     &  $\boldsymbol{4 \times 10^{2}}$    &  $\boldsymbol{80}$                 &  $\boldsymbol{44}$       &  $\boldsymbol{\propto k^{-4}}$  &  $\boldsymbol{0}$                   &  $\boldsymbol{17.17}$           &  $\boldsymbol{0.0017}$      &  $\boldsymbol{0.0022}$       &  $\boldsymbol{66}$      &  $\boldsymbol{0.00043}$  &  $\boldsymbol{0.054}$              &  $\boldsymbol{0.45}$               \\
$A80b$      &  $448^3$     &  $1\times 10^{-4}$     &  $1\times 10^{-4}$     &  $4 \times 10^{4}$    &  $80$                 &  $44$       &  $\propto k^{-4}$  &  0                   &  $17.17$           &  $0.0017$      &  $0.0022$       &  $66$      &  $0.00043$  &  $0.055$              &  $0.45$               \\
$A80c$      &  $432^3$     &  $1\times 10^{-4}$     &  $1\times 10^{-4}$     &  $4 \times 10^{6}$    &  $80$                 &  $41$       &  $\propto k^{-4}$  &  0                   &  $15.00$           &  $0.0015$      &  $0.0021$       &  $38$      &  $0.00014$  &  $0.0077$             &  $0.47$               \\
$A80d$      &  $432^3$     &  $1\times 10^{-4}$     &  $1\times 10^{-4}$     &  $4 \times 10^{10}$  &  $80$                 &  $41$       &  $\propto k^{-4}$  &  0                   &  $-$  &  $0.00018$     &  $0.0021$       &  $6$     &  $0.0001$     &  $0.0039$             &  $0.6$                \\
$A200$      &  $448^3$     &  $4 \times 10^{-4}$  &  $1\times 10^{-4}$     &  $4 \times 10^{2}$    &  $2 \times 10^{2}$    &  $44$       &  $\propto k^{-4}$  &  0                   &  $1.70$            &  $0.00068$     &  $0.0022$       &  $65$      &  $0.00044$  &  $0.023$              &  $2.2$                \\
$A400$      &  $448^3$     &  $4 \times 10^{-4}$  &  $1\times 10^{-4}$     &  $4 \times 10^{2}$    &  $4 \times 10^{2}$    &  $44$       &  $\propto k^{-4}$  &  0                   &  $0.85$            &  $0.00034$     &  $0.0022$       &  $74$      &  $0.00042$  &  $0.015$              &  $1.5$                \\
$\boldsymbol{A1000}$     &  $\boldsymbol{448^3}$     &  $\boldsymbol{4 \times 10^{-4}}$  &  $\boldsymbol{1\times 10^{-4}}$     &  $\boldsymbol{4 \times 10^{2}}$    &  $\boldsymbol{1\times 10^{3}}$      &  $\boldsymbol{44}$       &  $\boldsymbol{\propto k^{-4}}$  &  $\boldsymbol{0}$                   &  $\boldsymbol{0.35}$            &  $\boldsymbol{0.00014}$     &  $\boldsymbol{0.0022}$       &  $\boldsymbol{94}$      &  $\boldsymbol{0.00041}$  &  $\boldsymbol{0.0039}$             &  $\boldsymbol{2.2}$                \\
$B40a$      &  $720^3$     &  $1\times 10^{-4}$     &  $2 \times 10^{-4}$  &  $4$                  &  $40$                 &  $48$       &  $\propto k^{-4}$  &  0                   &  $84.77$           &  $0.0085$      &  $0.0048$       &  $45$      &  $0.0045$     &  $0.69$               &  $5$                  \\
$B40$       &  $720^3$     &  $1\times 10^{-4}$     &  $2 \times 10^{-4}$  &  $4 \times 10^{2}$    &  $40$                 &  $48$       &  $\propto k^{-4}$  &  0                   &  $84.70$           &  $0.0085$      &  $0.0048$       &  $45$      &  $0.0044$     &  $0.68$               &  $4.8$                \\
$B40b$      &  $720^3$     &  $1\times 10^{-4}$     &  $2 \times 10^{-4}$  &  $4 \times 10^{4}$    &  $40$                 &  $48$       &  $\propto k^{-4}$  &  0                   &  $84.70$           &  $0.0085$      &  $0.0048$       &  $45$      &  $0.0037$     &  $0.61$               &  $4$                  \\
$B40c$      &  $720^3$     &  $1\times 10^{-4}$     &  $2 \times 10^{-4}$  &  $4 \times 10^{6}$    &  $40$                 &  $48$       &  $\propto k^{-4}$  &  0                   &  $84.77$           &  $0.0085$      &  $0.0048$       &  $41$      &  $0.001$      &  $0.17$               &  $0.92$               \\
$B40d$      &  $720^3$     &  $1\times 10^{-4}$     &  $2 \times 10^{-4}$  &  $4 \times 10^{8}$    &  $40$                 &  $48$       &  $\propto k^{-4}$  &  0                   &  $85.15$           &  $0.0022$      &  $0.0048$       &  $30$      &  $0.00027$  &  $0.016$              &  $0.3$                \\
$C80c$      &  $720^3$     &  $1\times 10^{-4}$     &  $1\times 10^{-4}$     &  $4 \times 10^{6}$    &  $80$                 &  $96$       &  $\propto k^{-4}$  &  0                   &  $84.88$           &  $0.0085$      &  $0.0048$       &  $90$      &  $0.0015$     &  $0.24$               &  $1.2$                \\
$R$--$2$  &  $672^3$         &   
$0$ &
$2 \times 10^{-4}$  &  $4 \times 10^{2}$    &  $0$                  &   $-$           &  0                   &  $\propto k^{-2}$  &  $-$              &   $-$              &   $-$             &  $50$      &  $0.18$       &      
$140$  &  $460$  \\
\hline
\end{tabular}
\label{tab_DNSoverview}
\end{table*}

\subsection{Production of turbulence and mean-field dynamo instability}
\label{sec_LSD}

Magnetic fluctuations generated by the chiral dynamo instability, produce velocity fluctuations $U_\mathrm{rms}$ through the Lorentz force.
This leads to an increase of   
the Reynolds number $\mathrm{Re}_\mathrm{M} = U_\mathrm{rms}/(k_\mathrm{f} \eta)$,
where $k_\mathrm{f}$ is the forcing wave number.
In such magnetically driven turbulence, $k_\mathrm{f}$ is roughly 
equal to the wave number on which the magnetic energy peaks. 
For a chiral dynamo instability, this corresponds 
to $k_\mathrm{f}\approx k_\mathrm{5} = \mu_{5,\mathrm{max}}/2$.
Using the rough assumption that 
$\sqrt{\meanrho} \, U_\mathrm{rms} \approx B_\mathrm{rms}$,
we can estimate the critical magnetic field strength 
$B_\mathrm{crit}$ which is necessary for the production of turbulence, i.e. for reaching a value of $\mathrm{Re}_\mathrm{M}$ above unity.
We find that the critical magnetic field strength is estimated as
\begin{eqnarray}
    B_\mathrm{crit} \approx \sqrt{\meanrho} \, \eta 
    \frac{\mu_\mathrm{max}(t_0)}{2}.
\label{eq_Bcrit}
\end{eqnarray}
The value of $B_\mathrm{crit}$ is presented as horizontal black lines 
in Fig.~\ref{fig_Bsat_C5Cmu}. 
It can be used to illustrate the regions of the 
parameter regime in which turbulence can be produced.

If the small-scale chiral dynamo leads to a magnetic field
that exceeds $B_\mathrm{crit}$,
a mean-field dynamo instability can be excited.
The maximum growth rate of the mean-field dynamo is
\begin{eqnarray}
  \gamma_\alpha= \frac{(\eta \langle{\mu_{5}}\rangle + \alpha_\mathrm{M})^2}{4(\eta+\eta_\mathrm{T})},
\label{eq_gammaalpha}
\end{eqnarray}
where $\langle{\mu_{5}}\rangle$ 
is the mean chiral chemical potential and 
$\alpha_\mathrm{M}$ is the magnetic $\alpha$ effect 
\citep{SchoberEtAl2022a,SchoberEtAl2022b}.
Here $\alpha_\mathrm{M} = 2(q-1)/(q+1) \, 
\tau_{\rm c} \, \chi_{\rm c} / \meanrho$
is the magnetic $\alpha$ effect, which is determined by the current helicity
$\chi_{\rm c} = \bra{{\bm b} {\bm \cdot} ({\bm \nabla} \times{\bm b})} \approx \bra{ {\bm a}\cdot {\bm b}} \, k_\mathrm{f}^2$,  
where $q$ is the exponent of the magnetic energy 
spectrum $E_{\rm M}\propto k^{-q}$,
and ${\bm a}$ and ${\bm b}$ are the fluctuations of the vector potential and the magnetic field, respectively.
The correlation time of the magnetically-driven turbulence is 
$\tau_{\rm c} \approx (U_\mathrm{A} k_\mathrm{f})^{-1}$, 
where the Alfv\'en speed is 
$U_{\rm A}=\sqrt{\bra{{\bm b}^2}} /\sqrt{\meanrho} \approx B_\mathrm{rms}/\sqrt{\meanrho}$. 
The turbulent diffusion coefficient $\eta_\mathrm{T}$ is estimated as 
$\eta_\mathrm{T}=U_\mathrm{rms}/(3 k_\mathrm{f})$.
The characteristic wave number on which the mean-field dynamo occurs is 
\begin{eqnarray}
  k_\alpha= \frac{|\eta \langle{\mu_{5}}\rangle + \alpha_\mathrm{M}|}{2(\eta+\eta_\mathrm{T})}.
\label{eq_kalpha}
\end{eqnarray}

\begin{figure*}[t!]
\centering
  \includegraphics[width=\textwidth]{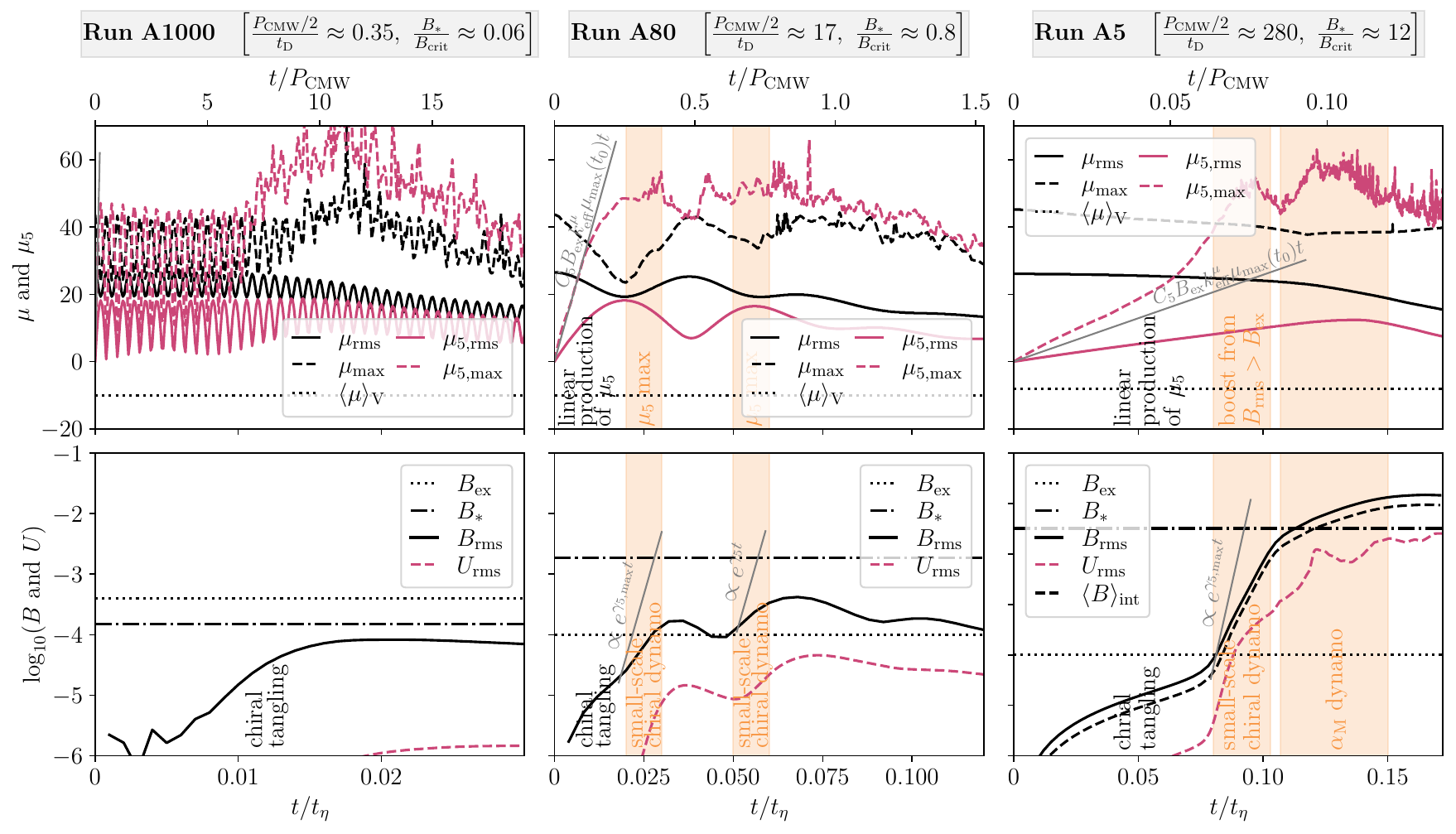} 
  \caption{Exemplary runs from the three different regimes: 
  A high-frequency CMW with just chiral tangling (Run $A1000$, \textit{left panels}),
  inefficient small-scale chiral dynamo due to a CMW with moderate frequency (Run $A80$, \textit{middle panels}), 
  and a
  low-frequency CMW with a small-scale chiral and mean-field dynamo (Run $A5$, \textit{right panels}). 
  The \textit{top panels} show the time evolution 
  of the rms and maximum values of the chemical potential $\mu$ 
  and the chiral chemical potential $\mu_5$, respectively,
  as well as the volume average of $\mu$.
  The \textit{bottom panels} show the evolution of the rms values 
  of the magnetic and the velocity fields, $B_\mathrm{rms}$ and  $U_\mathrm{rms}$
  as well as the external field strength $B_\mathrm{ex}$ and the maximum 
  possible magnetic field strength $B_*$ if no turbulence is produced.
  For Run $A5$ the time evolution of the mean magnetic field 
  strength $\langle B \rangle_\mathrm{int}$
  is presented for comparison.
  }
\label{fig_examples_ts}
\end{figure*}

\begin{figure*}[t]
\center
  \includegraphics[width=0.95\textwidth]{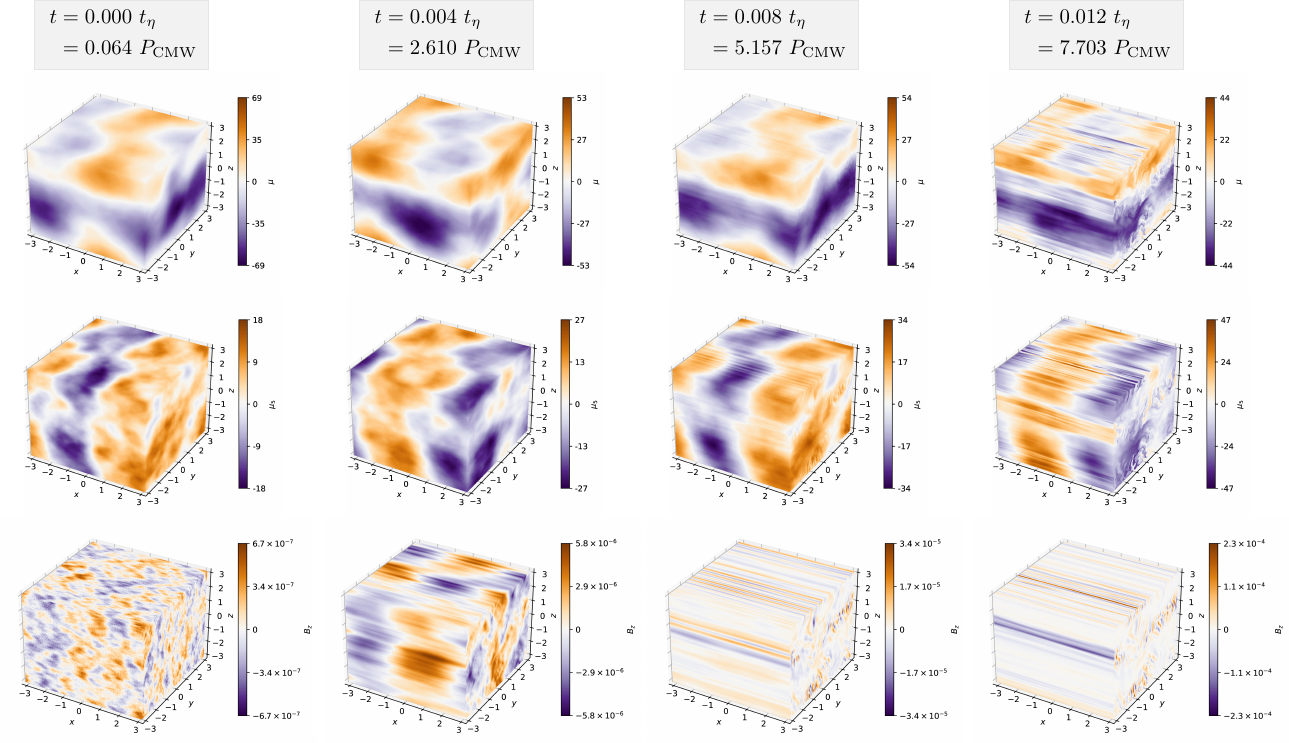} 
  \caption{Snapshots of Run $A1000$. 
  The surface of the cubic domain is shown for $\mu$ (upper row), 
  $\mu_5$ (middle row), and the $x$ component of the magnetic field 
  $B_z$ (lower row). 
  The snapshots cover the different evolutionary phases, 
  from the initial $\mu_5$ production phase ($t=0.05~t_\eta$, first column), 
  to the onset of the chiral dynamo instability ($t=0.10~t_\eta$, second column), 
  to the early mean-field dynamo stage ($t=0.15~t_\eta$, third column), 
  and the end of the mean-field dynamo stage ($t=0.20~t_\eta$, forth column).
  }
\label{fig_cubes_chiral_tangling}
\end{figure*}

\begin{figure*}[t]
\center
  \includegraphics[width=0.95\textwidth]{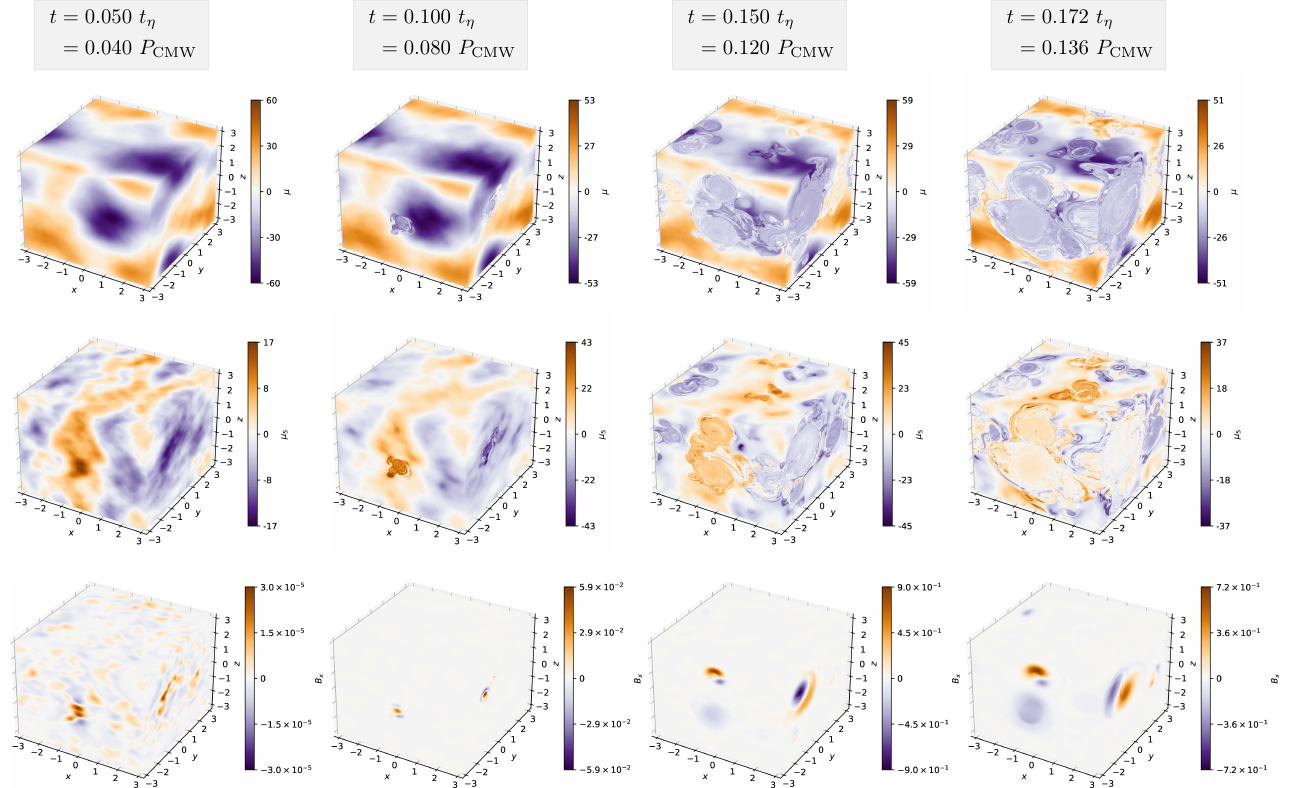}
  \caption{Similar to Fig.~\ref{fig_cubes_chiral_tangling} but for Run $A5$.
  }
\label{fig_cubes}
\end{figure*}

\begin{figure*}[t!]
\centering
  \includegraphics[width=\textwidth]{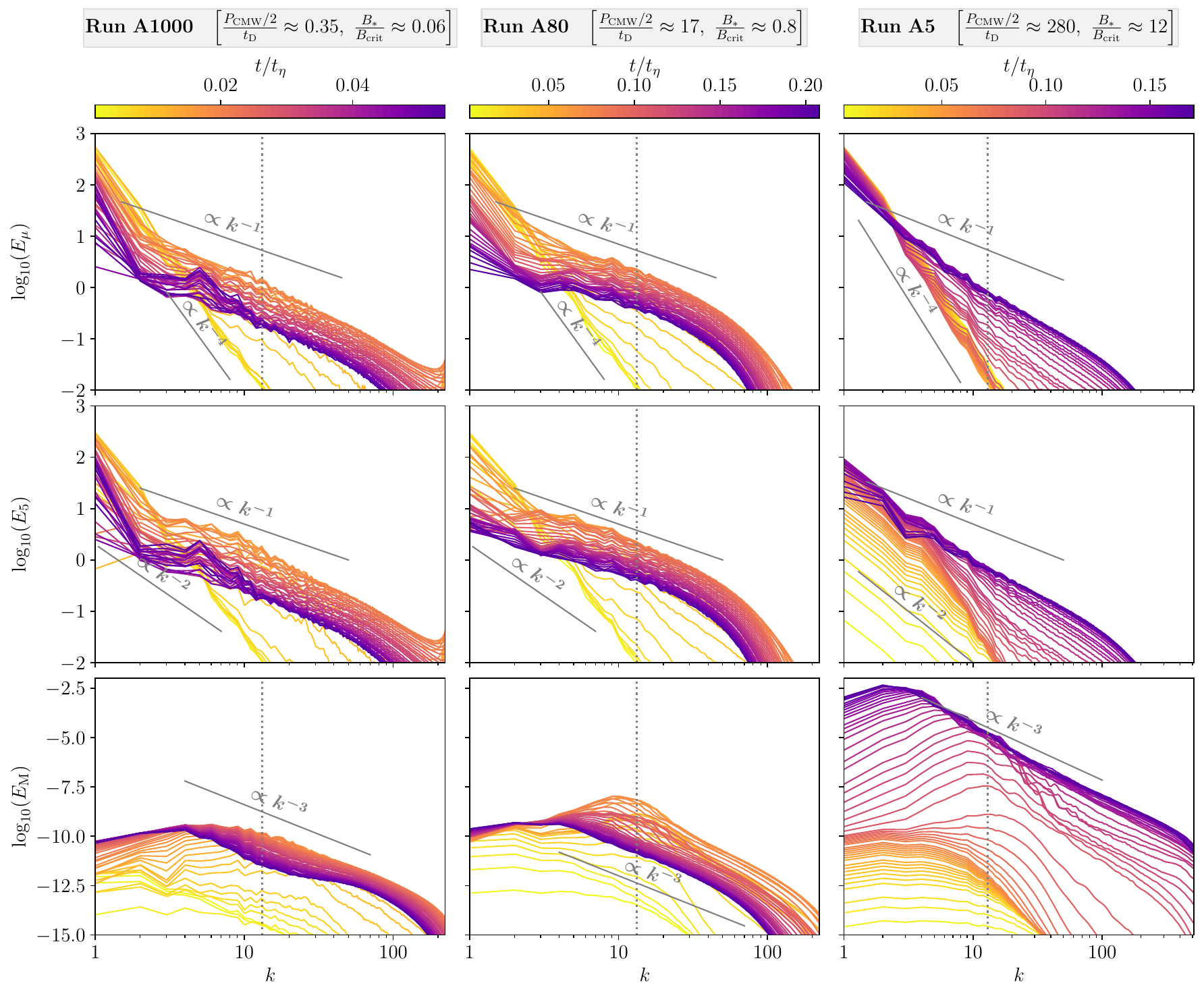} 
  \caption{Exemplary runs from the three different regimes: 
  A high-frequency CMW with just chiral tangling (Run $A1000$, \textit{left panels}),
  inefficient small-scale chiral dynamo due to a CMW with moderate frequency (Run $A80$, \textit{middle panels}), and a
  low-frequency CMW with small-scale chiral and mean-field dynamos (Run $A5$, \textit{right panels}). 
  From \textit{top} to \textit{bottom}, the spectrum of fluctuations of 
  chemical potential, $E_\mu(k)$, 
  chiral chemical potential, $E_5(k)$, and the magnetic energy spectrum $E_\mathrm{M}(k)$ are shown.
  The vertical dotted line indicates the highest possible value of the small-scale dynamo instability scale, $\mu_\mathrm{max}(t_0)/2$, 
  on which the magnetic field is amplified 
  if all of the initial $\mu$ has been converted to $\mu_5$.
  }
\label{fig_examples_spec}
\end{figure*}

\section{Numerical simulations}
\label{sec_DNS}

In this section, we use simulations to verify the phenomenology discussed above.
Using DNS, the conditions for chiral dynamo instabilities, efficient magnetic field
amplification and, in particular, the mean-field dynamo
phase can be analyzed qualitatively. 

\subsection{Setup and analysis tools}

We use the \textsc{Pencil Code} \citep{PencilCode2021} to
solve equations~(\ref{ind-DNS_CSE})--(\ref{mu5-DNS_CSE})
in a three-dimensional periodic domain of size $L^3 = (2\pi)^3$ with a resolution of
up to $1024^3$.
This code employs a third-order accurate time-stepping method \cite{Wil80}
and sixth-order explicit finite differences in space \citep{BD02,Bra03}.
The smallest wave number covered in the numerical domain 
is $k_1 = 2\pi/L = 1$ which we use for
normalization of length scales.
All velocities are normalized to the sound speed $\cs = 1$ and the
mean fluid density is set to $\meanrho = 1$.
Further, the magnetic Prandtl number is $1$,
i.e.\ the magnetic diffusivity equals the viscosity.
Time is normalized either by the diffusion time $t_\eta = (\eta k_1^2)^{-1}$
or by the period of the chiral magnetic wave $P_\mathrm{CMW}$.

The simulation parameters have been selected to cover 
the three different regimes: the ``chiral tangling regime'' ($t_\mathrm{D} \gtrsim P_\mathrm{CMW}/2$),
the ``small-scale chiral dynamo regime''
($t_\mathrm{D} \lesssim P_\mathrm{CMW}/2$ and $B_* \lesssim B_\mathrm{crit}$) and the
``mean-field dynamo regime''
($t_\mathrm{D} \lesssim P_\mathrm{CMW}/2$ and $B_* \gtrsim B_\mathrm{crit}$).
We also perform a comparison with the
results obtained in our previous study Ref.~\citep{SchoberEtAl2022a} (see Run $R$--$2$ there),
where chiral dynamo instabilities were found for an initial $\mu_5 \not = 0$ 
with zero mean but spatial fluctuations. 
For this comparison, the spatial maximum value of the chemical potential at the initial time $t_0$, $\mu_{\mathrm{max}}$, and its spectrum $E_\mu(k,t_0)$
have been chosen to eventually (before the onset of the 
small-scale chiral instability) result in a state of the system that is comparable to the initial conditions
in the Run $R$--$2$.
In particular, in the Run $R$--$2$ the initial $\mu_{5,\mathrm{max}}$ was $\approx50$ 
and the initial spectrum was 
$E_5(k,t_0)\propto k^{-2}$.
We therefore choose, for most runs of this study,
$\mu_{\mathrm{max}}(t_0) \approx50$ and $E_\mu(k,t_0)\propto k^{-4}$
which results in the spectrum $E_5(k,t_0)\propto k^{-2}$ according to
Eq.~(\ref{eq_mu5spec}).

The range of parameters chosen for this study is also
based on numerical aspects.
The parameter space that we explore includes the regime where the magnetic
field strength becomes larger than the critical value $B_\mathrm{crit}$
for the production of turbulence and the subsequent excitation of
mean-field dynamos.
According to the estimate in Eq.~(\ref{eq_BCMW}), which is illustrated in Fig.~\ref{fig_Bsat_C5Cmu}, 
the maximum magnetic field strength $B_*$ is higher for 
lower frequencies $\omega_\mathrm{CMW}$ of the CMW. 
However, for low $\omega_\mathrm{CMW}$ and therefore low values of $C_5$,
the initial linear (in time) production of $\mu_5$ becomes very slow, as can be
seen in Eq.~(\ref{eq_mu5_init_gen}).
Increasing the initial value of $\mu$ increases the initial production
rate of $\mu_5$, but this also leads to a larger value $B_*$, which can
cause the characteristic velocity of the CMW to become comparable or
larger than the sound speed.
Additionally, larger values of the initial $\mu$ lead to larger values of
$\mu_5$ and therefore a higher characteristic wave number of the chiral
dynamo instability.
Hence sufficient spatial resolution is required.
More details on the numerical criteria are given in
Appendix~\ref{sec_DNScriteria}.

Due to the numerical constraints discussed above, 
and also to allow for an appropriate comparison with
the DNSs presented in 
\citep{SchoberEtAl2022a,SchoberEtAl2022b},
we initiate most of the simulations
with $\mu_\mathrm{max}(t_0)\approx 50$ and use $\eta=10^{-4}$. 
We name this main series of simulations as Series A. 
Series B has $\mu_\mathrm{max}(t_0)\approx 50$ and $\eta=2 \times10^{-4}$ and Series C has  $\mu_\mathrm{max}(t_0)\approx 100$ and $\eta=10^{-4}$.
A summary of all runs of this study is given in 
Table~\ref{tab_DNSoverview} and the values for the 
corresponding estimates of $B_*$ is shown in Fig.~\ref{fig_Bsat_C5Cmu}.
Comparing the estimates $B_*$ and $B_\mathrm{crit}$, 
we can expect the occurrence of turbulence in 
Runs $A1$, $A3$, $A5$, $A5b$, and potentially in Runs $A10$ and $A10b$.
All other runs are expected to result in values of $B_*$ 
that are comparable or below $B_\mathrm{crit}$.

For runs in which turbulence develops, we perform 
a mean-field analysis. 
To this end, an averaging of the instantaneous fields needs to be performed in the DNS.
Since turbulence is driven magnetically, the forcing scale $k_f$ corresponds to the integral scale of the magnetic field which we determine via the magnetic 
energy spectrum $E_\mathrm{M}(k)$ as
\begin{eqnarray}
   k_\mathrm{int} \equiv \left[\frac{1}{{\cal E}_{\rm M}} \int
   E_\mathrm{M}(k)\,k^{-1}~\mathrm{d}k\right]^{-1} .
\label{eq_kint}
\end{eqnarray}
Magnetic energy density ${\cal E}_{\rm M}$ and magnetic spectrum
$E_\mathrm{M}(k)$ are connected as
\begin{eqnarray}
  {\cal E}_{\rm M} \equiv \frac{B_\mathrm{rms}^2}{2} =  \int E_\mathrm{M}(k)\,\mathrm{d}k.
\end{eqnarray}
To take into account that the magnetically driven 
turbulence exists in the range of the wave numbers $k \geq k_\mathrm{int}$,
we define the mean quantity $X$ in simulations as
\begin{eqnarray}
  \bra{X}_{\mathrm{int}} \equiv \left[\int E_\mathrm{X}(k) f(k)\,\mathrm{d}k\right]^{1/2},
\end{eqnarray}
where we use the function 
\begin{eqnarray}
    f(k)\equiv\left[1-\mathrm{tanh}(k-
     k_\mathrm{int})\right]/2
\label{eq_fk}
\end{eqnarray}
to filter out the scale $k \gtrsim k_\mathrm{int}$.
The result of taking the average $\bra{X}_{\mathrm{int}}$ is typically different from
the volume average, which is denoted by $\bra{X}_{\mathrm{V}}$.

\subsection{Results for the reference runs}

In this section, we present three reference runs
that have the same initial chemical potential,
but different frequencies of the CMW.
Run $A1000$ is the run in our sample with the highest
frequency of the CMW.
With a ratio of $P_\mathrm{CMW}/(2 t_\mathrm{D})\approx 0.35$, no dynamo activity is expected in Run $A1000$. 
The second reference run is $A80$, which has $P_\mathrm{CMW}/(2 t_\mathrm{D})\approx 17$.
Therefore, a small-scale chiral dynamo can occur. 
However since the expected maximum magnetic field strength $B_* \approx 0.0017$ 
is lower than the critical value $B_\mathrm{crit} \approx 0.0022$
that is necessary for the production of turbulence, 
no mean-field dynamo is expected in Run $A80$.
A mean-field dynamo
can occur in the third reference run, Run $A5$, which has $B_*/B_\mathrm{crit} \approx 12$.
Run $A5$ is the run with the third to the highest value of $P_\mathrm{CMW}/(2 t_\mathrm{D})$ in our sample.
We discuss the results of the reference runs
in the following and confront them with the estimates 
based on the phenomenological estimates 
presented in Sec.~\ref{sec_phenomenology}.

The left panels of Fig.~\ref{fig_examples_ts} show the 
time evolution of various parameters of Run $A1000$.
In the top left panel, the oscillatory behavior of 
$\mu_\mathrm{rms}$ and $\mu_{5,\mathrm{rms}}$ is clearly seen and the time evolution
governs almost $20$ periods of the CMW.
In systems like this, where the CMW has a very high frequency and the
initial chiral chemical potential is small, the timescale of 
the chiral dynamo $t_\mathrm{D}$ is much longer than $P_\mathrm{CMW}$.
In this case, magnetic fluctuations can only be amplified by chiral
tangling.
This phenomenon alone leads to the production of magnetic fluctuations
that are of the order of the imposed magnetic field $B_\mathrm{ex}$.
The magnetic field evolution in Run $A1000$
can be seen in the lower left panel of Fig.~\ref{fig_examples_ts}.
The maximum value of $B_\mathrm{rms}$ produced by chiral tangling alone is approximately less than half of $B_\mathrm{ex}$.
At $t\gtrsim 0.02~t_\eta$, both $\mu_\mathrm{rms}$ and $\mu_{5,\mathrm{rms}}$ decay, and therefore $B_\mathrm{rms}$ decreases.

Snapshots of Run $A1000$ are presented in 
Fig.~\ref{fig_cubes_chiral_tangling}.
While the magnetic field is, as in all simulations of this paper, set
up as weak and random fluctuations, the magnetic fluctuations quickly
develop into patches that are stretched along the $x$ axis.
At the forth snapshot shown here (at $t \approx 7.7~P_\mathrm{CMW}$), the $B_x$ 
patches stretch out through the entire numerical domain. 
The magnetic field structure produced by chiral tangling is therefore
very different from what is expected when a small-scale chiral dynamo
instability is excited.
In linear theory, the magnetic field instability 
is expected to occur on a characteristic wave number that 
is half of the value of $\mu_{5,\mathrm{max}}$.
This leads to the formation of isotropic patches of high absolute values
of $B_x$ on the surface of the numerical domain, at the locations where
$\mu_{5}$ reaches the maximum value.

The middle panels of Fig.~\ref{fig_examples_ts} show the 
time evolution of various parameters of Run $A80$, 
where a small-scale chiral dynamo instability occurs. 
Two oscillations between $\mu_\mathrm{rms}$ and 
$\mu_{5,\mathrm{rms}}$ are seen in the upper middle panel. 
The initial production of the spatial maximum value of $\mu_5(t)$,
$\mu_{5,\mathrm{max}}(t)$, proceeds linear in time and follows the
prediction given by Eq.~(\ref{eq_mu5_init_gen}) until the instant
$t\approx 0.015\, t_\eta$.
Temporal maxima of $\mu_{5,\mathrm{rms}}$ are reached at 
$t\approx 0.025 \, t_\eta$ and $t\approx 0.055\, t_\eta$.
These times coincide, as expected, with an increased 
growth rate of the magnetic field; see the time evolution
of magnetic fluctuations in the lower middle panel. 
However, the magnetic field fluctuations, $B_\mathrm{rms}$, never reach a field strength that is much larger than the one of the imposed field $B_\mathrm{ex}$. 
At its maximum, the rms magnetic field strength is 
approximately $0.2~B_*$ in Run $A80$.
At later times, $t\gtrsim 1.0\, P_\mathrm{CMW}$, the quantities
$\mu_{\mathrm{rms}}$, $\mu_{5,\mathrm{rms}}$, and $B_\mathrm{rms}$ decay.

The estimated maximum value of $B_\mathrm{rms}$, $B_*$, is $16$ times
higher in Run $A5$ than in $A80$.
Contrary to the other reference runs, in Run $A5$, the maximum value
$B_*$ of magnetic field exceeds $B_\mathrm{crit}$, which implies that
turbulence can be produced.
The time evolution of Run $A5$ is shown in the right
panels of Fig.~\ref{fig_examples_ts}. 
Due to the smaller value of $\omega_\mathrm{CMW}$, the production of
$\mu_5$ is much slower than that in Run $A80$.
Here, the threshold for the small-scale chiral dynamo
instability is only being exceeded at $t\approx 0.08\, t_\eta$.
After the magnetic field has been amplified by more than two orders of
magnitude through the small-scale chiral dynamo, a mean-field dynamo
instability is excited at $t \approx 0.11~t_\eta$ with a growth rate
of the magnetic field that is less than that for the small-scale chiral
dynamo instability.
In Run $A5$, the magnetic field strength exceeds $B_*$ 
by a factor of $\approx 5.4$.
The analysis of the mean-field dynamo phase of Run $A5$ and the other
runs in which the value of the Reynolds number eventually exceeds unity
will be discussed in more detail in Sec.~\ref{sec_mean_field}.

The time evolution of the simulation snapshots for Run
$A5$ is presented in Fig.~\ref{fig_cubes}.
Here, the values for the quantities $\mu$, $\mu_5$, 
and the $x$ component of the magnetic field, $B_x$, 
on the surface of the cubic domain  
are shown for $t=0.05~t_\eta - 0.172~t_\eta$. 
The snapshots show that, as expected, $\mu_5$ grows 
fastest where the gradient of $\mu$ is largest.
At $t=0.05~t_\eta$, the fastest production of $\mu_5$ occurs 
approximately in the middle of the front $x$-$z$ plane (where $\mu_5$ is 
produced with a positive sign) and in the middle of the front of the 
$x$-$y$ plane (where $\mu_5$ is produced with a negative sign).
These are the two locations on the shown surface of the domain, where also 
the magnetic field instability kicks in the fastest. 
In the snapshot at time $t=0.10~t_\eta$, the magnetic field grows approximately on the length scale $k_5^{-1} \approx (\mu_5/2)^{-1} \approx 1/20$. 
At $t=0.20~t_\eta$, the simulation is at the end of the
mean-field dynamo stage and the characteristic length scale of the magnetic field has increased. 
At late times, we also observe that both $\mu$ and $\mu_5$ develop small-scale fluctuations, especially in locations where the magnetic field is the strongest.
These small-scale structures are symmetric in $\mu$ and 
$\mu_5$, but with opposite sign. 

For a quantitative analysis of the evolution of the characteristic scales, the evolution of 
the energy spectra is presented in Fig.~\ref{fig_examples_spec} for 
Runs $A1000$ (middle panels), $A80$ (middle panels), and $A5$ (right panels).
In all cases, the initial spectrum of $\mu_5$, $E_5(k)$, scales with
the wave number as $k^{-2}$, as expected for an initial $E_\mu(k)$
spectrum that is proportional to $k^{-4}$; see Eq.~(\ref{eq_mu5spec}).
For Runs~$A1000$ and $A80$, 
the initial $k^{-2}$ scaling of $E_5(k)$ is less visible
due to the fast production of $\mu_5$.
At later times, the spectra $E_5(k)$ and $E_\mu(k)$, approach a scaling
of $k^{-1}$, as has been reported in \citep{SchoberEtAl2022a}.
The evolution of the magnetic energy spectra $E_\mathrm{M}(k)$ is shown
in the lowest panels of Fig.~\ref{fig_examples_spec}.
In the case of Run $A80$, a short phase of amplification
on $k=\mu_\mathrm{max}(t_0)/2\approx22$ is seen, but at $t\gtrsim 0.07\, t_\eta$
the magnetic energy decays and a $E_\mathrm{M}\propto k^{-3}$ develops.
The magnetic field amplification is much more efficient in Run $A5$. 
Here the initial instability occurs also on $k=\mu(t_0)\approx22$.
Due to the production of turbulence, however, the peak of the magnetic
energy spectrum moves to smaller wave numbers.
Eventually, a $E_\mathrm{M}\propto k^{-3}$ develops in Run $A5$ as well.

\begin{figure}[t!]
\centering
\includegraphics[width=0.5\textwidth]{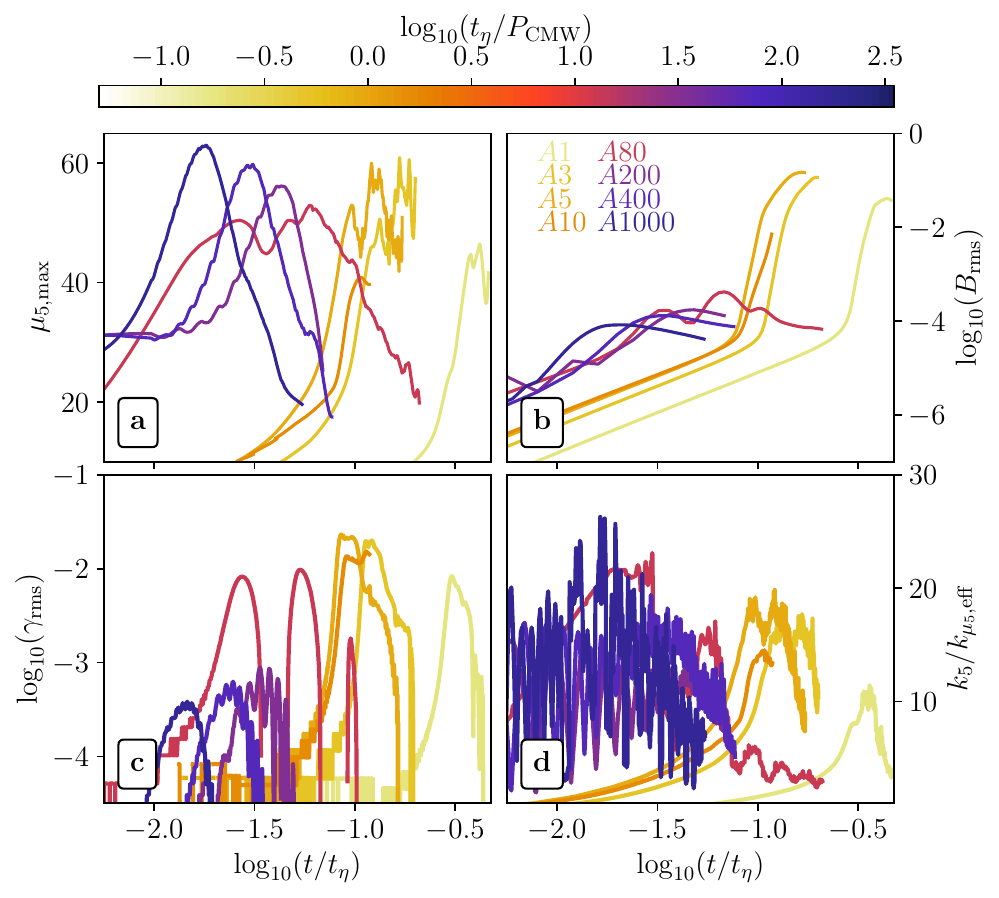}
  \caption{Time series for runs with different values of
  the ratio of the CMW period over the resistive time, $P_\mathrm{CMW}/t_\eta$,
  as indicated by the color bar. 
  \textit{(a)} Maximum values of chemical and chiral chemical potential, $\mu_{5,\mathrm{max}}$ and $\mu_{\mathrm{max}}$.
  \textit{(b)} Rms value of the magnetic field strength $B_\mathrm{rms}$.
  \textit{(c)} Measured growth rate $\gamma_\mathrm{rms}$ of $B_\mathrm{rms}$. 
  \textit{(d)} The measured scale separation $k_5/k_{\mu_5, \mathrm{eff}}$.
  }
  \label{fig_ts_m2_omCMW}
\end{figure}
\begin{figure}[t!]
\centering
\includegraphics[width=0.5\textwidth]{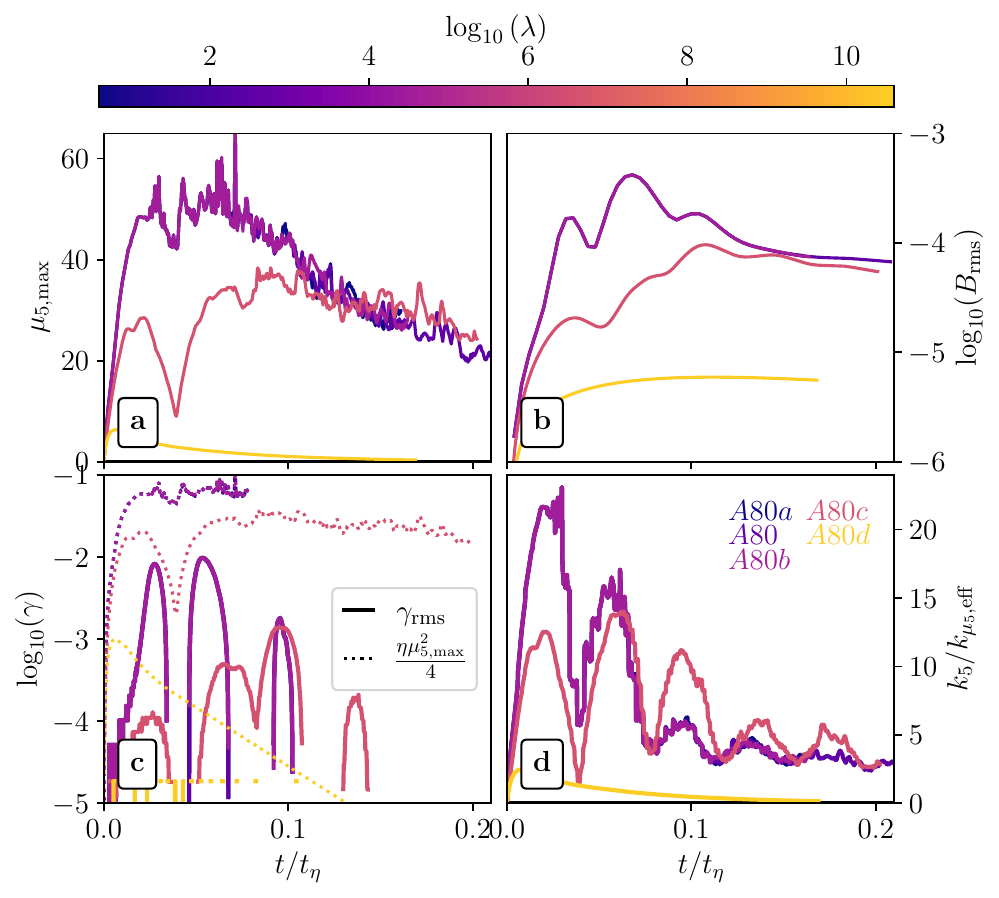}
  \caption{Time series for runs with different values of
  $\lambda$,
  as indicated by the color bar. 
  \textit{(a)} Maximum values of chemical and chiral chemical potential, $\mu_{5,\mathrm{max}}$ and $\mu_{\mathrm{max}}$.
  \textit{(b)} Rms value of the magnetic field strength $B_\mathrm{rms}$.
  \textit{(c)} Measured growth rate $\gamma_\mathrm{rms}$ of $B_\mathrm{rms}$, and
  the theoretical value $\eta \mu_{5,\mathrm{max}}^2/4$ of the maximum
  growth rate, which would be reached for an infinite scale separation
  ($k_5 \gg k_{\mu_5, \mathrm{eff}}$).
  \textit{(d)} The measured scale separation $k_5/k_{\mu_5, \mathrm{eff}}$.
  }
  \label{fig_ts_m2_lambda}
\end{figure}
\begin{figure}[t!]
\centering
\includegraphics[width=0.5\textwidth]{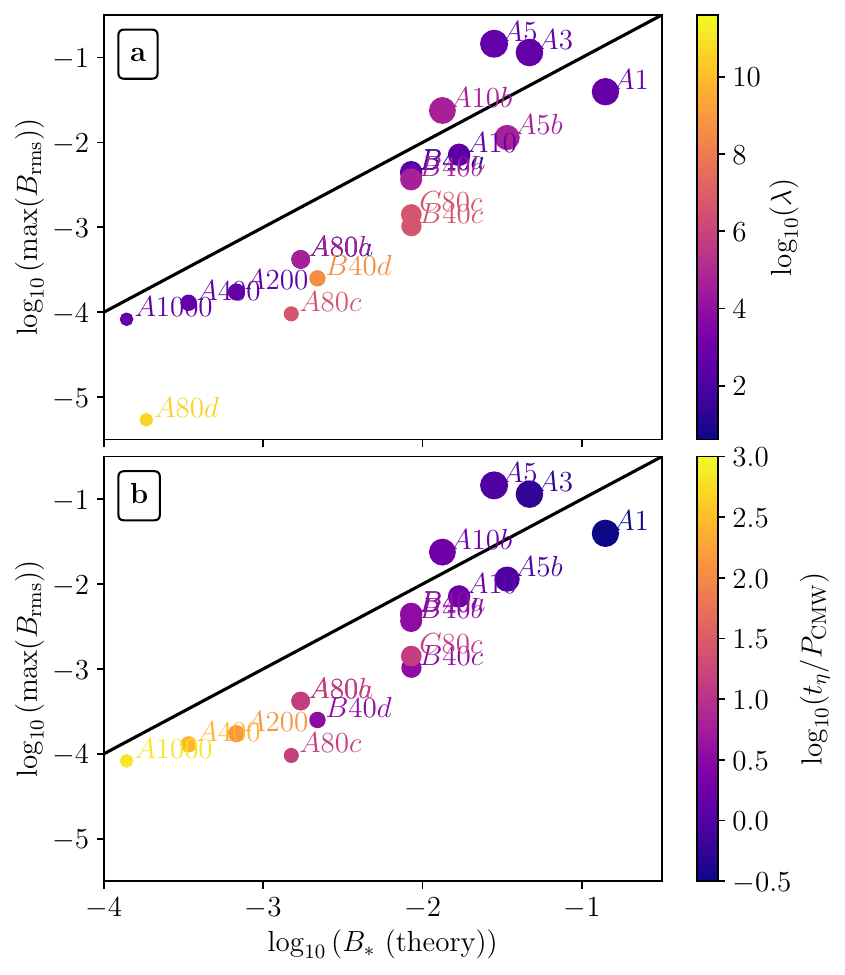}
  \caption{The maximum value of $B_\mathrm{rms}$ measured in 
  all runs as a function of the phenomenologically derived 
  maximum, $B_*$.
  Colors indicate \textit{(a)} the value of $\lambda$ and 
  \textit{(b)} the ratio of the resistive time $t_\eta$ over the period $P_\mathrm{CMW}$ of the CMW   
  [note that Run $A80d$ is not shown in panel \textit{b} 
  since there is no CMW].
  The size of the symbols increases with increasing maximum 
  Reynolds number obtained in the individual runs.
  }
  \label{fig_Bmax_Bmaxtheo}
\end{figure}

\subsection{Exploration of the parameter space}

A direct comparison between runs with different CMW frequencies
$\omega_\mathrm{CMW}$, including Runs $A5$,  $A80$, and $A1000$, is presented in
Fig.~\ref{fig_ts_m2_omCMW}.
All of the runs in Fig.~\ref{fig_ts_m2_omCMW} have 
the same initial values of $\mu$,
and the same $\lambda$ and $\eta$.
Even though the temporal maximum values of 
$\mu_{5,\mathrm{max}}$ is higher for runs with
higher $\omega_\mathrm{CMW}$, the maximum value of the 
produced magnetic field decreases with increasing 
$\omega_\mathrm{CMW}$.
Therefore, CMWs with higher frequencies are less efficient in amplifying the magnetic field. 
This stems from the small-scale chiral dynamo being
less efficient when the period of the CMW is small.

Figure~\ref{fig_ts_m2_lambda} shows a comparison between runs with
different chiral feedback parameter, $\lambda$.
As expected from Eq.~(\ref{eq_BCMW}), larger values of $\lambda$ lead
to lower magnetic field strengths.
In Fig~\ref{fig_ts_m2_lambda}a, it can be seen that $\mu_{5,\mathrm{max}}$ in 
all runs with low $\lambda$ reach a value that is comparable to (or even slightly exceeds)
the initial value of $\mu_{\mathrm{max}}$.
This leads to three instances of magnetic field amplification, see Fig~\ref{fig_ts_m2_lambda}b.
In Run $A80c$, which has $\lambda=4\times10^{6}$, lower values of $\mu_{\mathrm{max}}$
are reached, which is caused by the damping of the CMW according to Eq.~(\ref{damp_CMW}). 
However, phases of magnetic field amplification can still be seen for Run~$A80c$.
This is different for Run~$A80d$, which has $\lambda=4\times10^{8}$. 
Here, no CMW occurs since the frequency of the wave is imaginary. 

The maximum magnetic field strength found in DNS agrees well with
the prediction given by Eq.~(\ref{eq_BCMW}), as is illustrated in
Fig.~\ref{fig_Bmax_Bmaxtheo}.
Here, the predicted value $B_*$ is
plotted against the temporal maximum of the magnetic field strength in
all DNS of this study.
The agreement between phenomenology and DNS is better 
for runs with lower values of $\lambda$.
This follows from the fact that Eq.~(\ref{eq_BCMW}) is based on two
assumptions:
(i) the effective correlation wave number of $\mu$, 
$k_{\mu,\mathrm{eff}}$, stays constant until the
maximum magnetic field strength is reached, 
and (ii) $\mu_{5,\mathrm{max}}$ can 
reach the same value as $\mu_{\mathrm{max}}(t_0)$. 
As can be seen in Fig~\ref{fig_ts_m2_lambda}a, the conversion between
$\mu$ and $\mu_5$ becomes less efficient when $\lambda$ increases.
Even though all runs in Fig.~\ref{fig_ts_m2_lambda} have similar values
of $\mu_\mathrm{max}(t_0)\approx 45$, in the run $A80d$ the temporal
maximum of $\mu_\mathrm{5,max}$ never exceeds $6.3$.
Therefore, in the limit of large $\lambda$, the expression in Eq.~(\ref{eq_BCMW}) 
has to be considered as an upper limit 
for the maximum possible magnetic field strength.

\begin{figure*}[t!]
\centering
  \includegraphics[width=\textwidth]{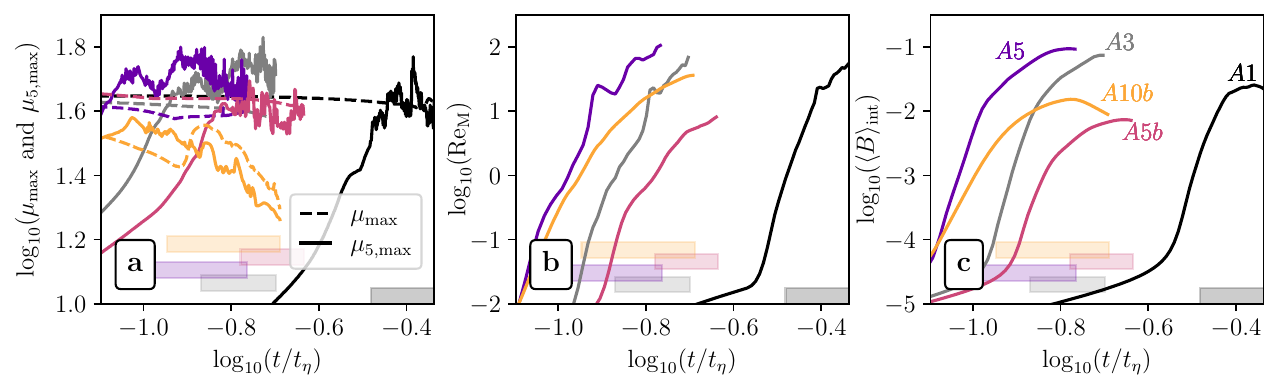} 
   \caption{
Time evolution of all runs in which magnetically dominated turbulence is produced.
Different colored lines indicate the different runs 
(A1, A3, A5, A5b, and A10b) as indicated in panel (c).
\textit{(a)} Time evolution of $\mu_\mathrm{max}$ (dashed lines) and 
$\mu_{5,\mathrm{max}}$ (solid lines).
\textit{(b)} Time evolution of the magnetic Reynolds number $\mathrm{Re}_\mathrm{M}$.
\textit{(c)} Time evolution of the magnetic field strength
$\langle B \rangle_\mathrm{int}$ averaged on the integral scale. 
The color bars in the panels highlight the time
range in which the Reynolds number is larger than unity. 
This time range is shown in Fig.~\ref{fig_ts_turb_all_meanfield}.
}
\label{fig_ts_turb_all}
\end{figure*}
\begin{figure*}[t!]
\centering
  \includegraphics[width=\textwidth]{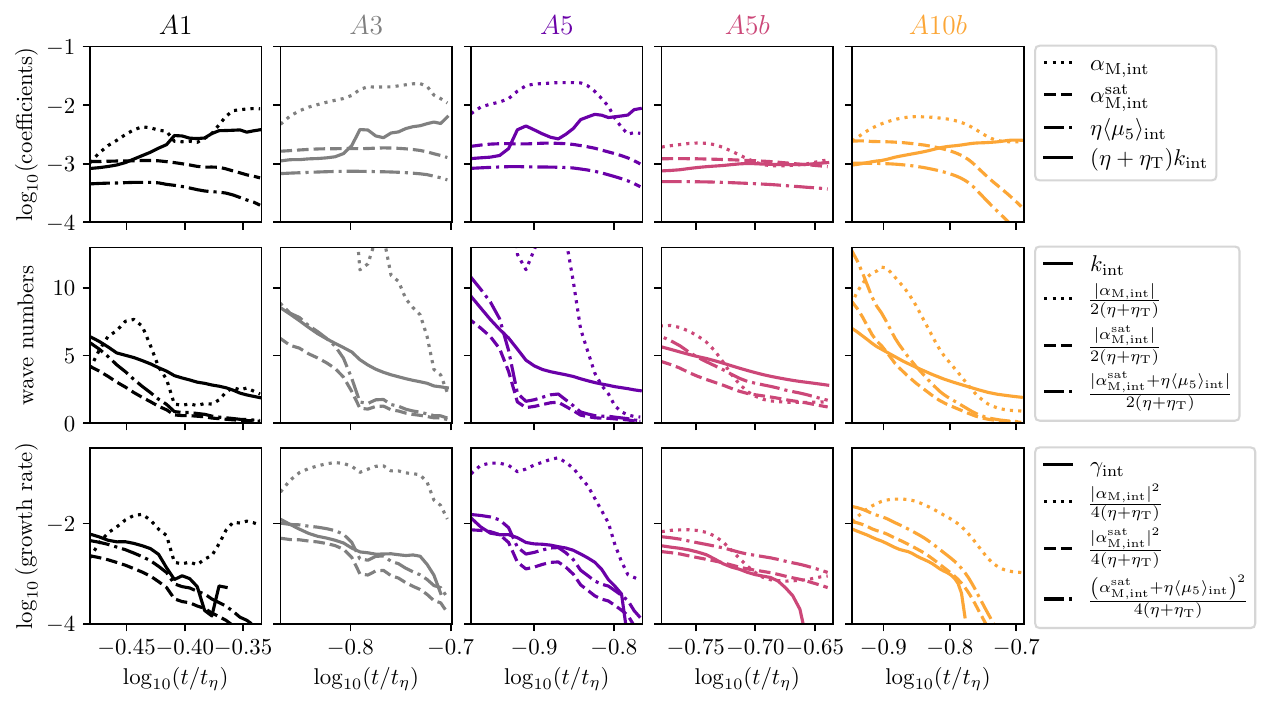} 
  \caption{Mean-field dynamo analysis for all runs in which 
  magnetically dominated turbulence is produced.
  Different rows of panels indicate the different runs 
  (A1, A3, A5, A5b, and A10b) as indicated on the top.
  The time axis is reduced to the phase of the DNS, where the Reynolds
  number is larger than unity up to the final time of the individual
  simulations.
  This time range is also indicated by the color blocks in panels (a)-(c) of Fig.~\ref{fig_ts_turb_all}.
  \textit{(Upper row)} Time evolution of dynamo 
   coefficients, $\alpha_\mathrm{M, int}$ (dotted lines), $\alpha^\mathrm{sat}_\mathrm{M, int}$ 
   (dashed lines), and $\eta + \eta_\mathrm{T}$
   multiplied by the integral 
   wave number $k_\mathrm{int}$ (solid lines). 
   The evolution of $\eta \langle\mu_5\rangle_\mathrm{M, int}$ (dashed-dotted lines) is also shown,
   which can be relevant for the mean-field dynamo.
   \textit{(Middle row)}
   Time evolution of the integral scale of the magnetic field 
   $k_\mathrm{int}$ as measured in the DNS (solid lines).
   This evolution is compared to theoretical estimates of the 
   mean-field dynamo theory that is based on the turbulent transport coefficients. 
   Different expressions for the magnetic $\alpha$ effect are used: 
   $\alpha_\mathrm{M, int}$ (dotted lines) and $\alpha^\mathrm{sat}_\mathrm{M, int}$ 
   (dashed lines and dashed-dotted lines).
   \textit{(Bottom row)} Same as middle row, but for the measured growth rate of $\langle B \rangle_\mathrm{int}$ in DNS, 
   $\gamma_\mathrm{int}$, and different theoretical estimates.}
\label{fig_ts_turb_all_meanfield}
\end{figure*}

\subsection{DNS with mean-field dynamo activity}
\label{sec_mean_field}

As can be seen in Fig.~\ref{fig_Bmax_Bmaxtheo}, the
maximum magnetic field tends to exceed the estimate from 
Eq.~(\ref{eq_BCMW}) for runs that develop turbulence.
This is understandable, since during the mean-field dynamo phase, the
sign of $\mu_5$ does not affect the magnetic field amplification.
Therefore, the comparison between the timescale of the CMW and
the chiral dynamo instability that leads to the estimate given by
Eq.~(\ref{eq_BCMW}), is not applicable in the presence of turbulence.
The magnetic field can grow to higher strengths, 
until saturation occurs due to 
nonlinear effects or the increase of turbulent magnetic diffusion.

In runs with low-frequency CMWs, the magnetic field strength reaches the
highest values, leading to efficient driving of magnetically dominated
turbulence.
In this case, a large-scale magnetic field is generated via a mean-field
dynamo instability, as can be seen in the snapshots of Run $A5$ in
Fig.~\ref{fig_cubes}.
Out of all the runs presented here, the ones in which 
$\mathrm{Re}_\mathrm{M}$ exceeds unity, i.e.~in which 
turbulence develops, are Runs $A1$, $A3$, $A5$, $A5b$, and $A10b$. 
The time series of various quantities of these runs 
are directly compared in Fig.~\ref{fig_ts_turb_all}. 
After the turbulence production phase, 
the value of $\mu_{5,\mathrm{max}}$ 
is comparable in Runs $A1$, $A3$, $A5$, and $A5b$. 
In Run~$A10b$, $\mu_{5,\mathrm{max}}$ never exceeds $40$, which is due
to the higher frequency of the CMW.
In all runs, the magnetic Reynolds number exceeds unity after
less than a resistive time; see Fig.~\ref{fig_ts_turb_all}b.
With $\mathrm{Re}_\mathrm{M}$ becoming larger than one, the type
of dynamo instability changes from a small-scale chiral dynamo to a
mean-field dynamo.
This transition, which is accompanied by a change in the growth rate,
can be seen in Fig.~\ref{fig_ts_turb_all}c, where the time evolution of
the mean magnetic field strength is presented.

The theoretically expected growth rate during the mean-field dynamo
phase is given by Eq.~(\ref{eq_gammaalpha}).
In the simulations, we estimate the magnetic $\alpha$ effect as 
$\alpha_\mathrm{M,int} = 2(q-1)/(q+1) \, \tau_{\rm c} \, \chi_{\rm c} \approx \tau_{\rm c} \bra{ {\bm a}\cdot {\bm b}}_\mathrm{int} \, k_\mathrm{int}^2$,  
assuming that the forcing scale is $k_\mathrm{f}\approx k_\mathrm{int}$ and that the exponent of the magnetic energy spectrum $q\approx3$.
The correlation time of the magnetically-driven
turbulence is $\tau_{\rm c} \approx (U_\mathrm{A} k_\mathrm{int})^{-1}$, where the Alfv\'en speed is
$U_{\rm A}=\sqrt{\bra{{\bm b}^2}}
\approx B_\mathrm{rms}$. 
The mean fluid density $\meanrho$ entering in $U_{\rm A}$ and $\alpha_\mathrm{M}$ is set to unity in the DNS.
The turbulent diffusion coefficient $\eta_\mathrm{T}$ is estimated as $\eta_\mathrm{T}=U_\mathrm{rms}/(3 k_\mathrm{int})$.
The time evolution of $\alpha_\mathrm{M,int}$ and 
$\eta_\mathrm{T}$ for all turbulent runs is presented in 
the upper panels of Fig.~\ref{fig_ts_turb_all_meanfield}.
The time range shown in Fig.~\ref{fig_ts_turb_all_meanfield}
is the moment when $\mathrm{Re}_\mathrm{M}$ exceeds unity up to the final time of the simulation, i.e.~it governs the turbulent phase of the simulation.
Right after the onset of turbulence, $\alpha_\mathrm{M,int}$ 
is the dominant transport coefficient for all runs presented in Fig.~\ref{fig_ts_turb_all_meanfield}.
However, $\eta_\mathrm{T}$ grows constantly with time.

The magnetic $\alpha$ effect can also be estimated from the evolutionary
equation for the magnetic helicity $\bra{ {\bm a}\cdot {\bm b}}$ of
the small-scale field ${\bm b}={\bm \nabla} \times{\bm a}$ in chiral
MHD \citep{RogachevskiiEtAl2017}:
\begin{eqnarray}
{\partial \over \partial t}  \bra{{\bm a} {\bm \cdot} {\bm b}} + 
\nabla \cdot {\bm F}
&=& 2 \eta \, \meanmufive \bra{{\bm b}^2} - 2 \meanEMF \cdot \meanBB 
\nonumber \\ 
&& - 2 \eta \, \bra{{\bm b} \, ({\bm \nabla} \times {\bm b})}, 
\label{MH1}
\end{eqnarray}
where ${\bm F}$ is the flux of $\bra{{\bm a} {\bm \cdot} {\bm b}}$
that is given by
\begin{eqnarray}
{\bm F} &=&  \left\langle {\bm u} \,a_j \right\rangle
\left\langle B_j \right\rangle - \meanBB \left\langle {\bm a} \cdot{\bm u} \right\rangle
- \eta \left\langle {\bm a} \times ({\bm \nabla}
\times{\bm b}) \right\rangle 
\nonumber\\
&& +\left\langle {\bm a} \times ({\bm u}
\times{\bm b}) \right\rangle ,
\label{PP4}
\end{eqnarray}
and $\meanEMF \equiv \bra{{\bm u} {\bm \times} {\bm b}}=\alpha_\mathrm{M} \meanBB - 
\eta_\mathrm{T} \, ({\bm \nabla} \times \meanBB)$
is the turbulent electromotive force.
In the steady-state, two leading source/sink
terms in Eq.~(\ref{MH1}),
$2 \eta \, \meanmufive  \bra{{\bm b}^2} - 2 \alpha_\mathrm{M}  \meanBB^2$,
compensate each other, so that
the magnetic $\alpha$ effect reaches \citep{SchoberEtAl2022a,SchoberEtAl2022b}
\begin{eqnarray}
  \alpha_\mathrm{M}^{\rm sat} = \eta \, \meanmufive \, \frac{\bra{{\bm b}^2}}{\meanBB^2}.
\label{eq_alphasat}
\end{eqnarray}
The time evolution of $\alpha_\mathrm{M, int}^{\rm sat} = \eta \bra{\mu_5}_\mathrm{int} B_\mathrm{rms}^2/\bra{B}_{\mathrm{int}}^2$ 
is compared to the one of $\alpha_\mathrm{M,int}$ 
in the upper panels of Fig.~\ref{fig_ts_turb_all_meanfield}.
We note that the values of $\alpha_\mathrm{M, int}^{\rm sat}$ are
consistently lower than $\alpha_\mathrm{M, int}$, which could result
from the fact that the divergence of the magnetic helicity fluxes is
ignored in the estimate of $\alpha_\mathrm{M, int}^{\rm sat}$.

In the middle and lower panels of Fig.~\ref{fig_ts_turb_all_meanfield}, 
the estimates of the turbulent transport coefficients are used 
to calculate the theoretically expected characteristic 
wave number $k_\alpha$ and growth rate $\gamma_\alpha$  
of the mean-field dynamo, respectively. 
The theoretical estimates, 
given by
Eqs.~(\ref{eq_gammaalpha}) and (\ref{eq_kalpha}), are compared with the 
measured characteristic 
wave number of the magnetic field, $k_\mathrm{int}$ and 
the measured growth rate $\gamma_\mathrm{int}$ of the 
mean magnetic field strength $\langle B \rangle_\mathrm{int}$.
In the middle row of Fig.~\ref{fig_ts_turb_all_meanfield} we compare the measured $k_\mathrm{int}$ to $|\alpha_\mathrm{M,int}|/\left[2(\eta + \eta_\mathrm{T}))\right]$ and $|\alpha_\mathrm{M,int}^\mathrm{sat}|/\left[2(\eta + \eta_\mathrm{T})\right]$, respectively,
and in the bottom row, we compare the measured $\gamma_\mathrm{int}$
to $|\alpha_\mathrm{M,int}|^2/\left[4(\eta + \eta_\mathrm{T})\right]$ and $|\alpha_\mathrm{M,int}^\mathrm{sat}|^2/\left[4(\eta + \eta_\mathrm{T})\right]$, respectively.
Using the $\alpha_\mathrm{M, int}$ to estimate $k_\alpha$ and 
$\gamma_\alpha$, tends to lead to slightly higher values than the measured 
$k_\mathrm{int}$ and $\gamma_\mathrm{int}$, 
while using $\alpha_\mathrm{M, int}^{\rm sat}$, 
leads to slightly lower values. 

One issue that arises in the comparison with theory is that while computing the mean value of 
$\langle \mu_5 \rangle_\mathrm{int}$ 
the information about the sign is lost, as the averaging process is based on the spectrum of $\mu_5^2$.
This is a problem because the expressions of $k_\alpha$ and $\gamma_\alpha$, as given in Eqs.~(\ref{eq_kalpha}) and (\ref{eq_gammaalpha}), include the sum of 
$\eta \langle{\mu_{5}}\rangle$ and $\alpha_\mathrm{M}$.
For strong turbulence, we expect that $\alpha_\mathrm{M} \gg \eta \langle{\mu_{5}}\rangle$, and therefore we neglect the $\eta \langle{\mu_{5}}\rangle_\mathrm{int}$ term in the estimates.
But for systems with low Reynolds numbers, the sign of $\langle \mu_5 \rangle_\mathrm{int}$ 
can be relevant in the comparison between DNS and mean-field theory.
As can be seen in the upper row of Fig.~\ref{fig_ts_turb_all_meanfield}, indeed,
in our simulations, the contribution of 
$\eta \langle{\mu_{5}}\rangle_\mathrm{int}$ can be relevant as it is not much smaller than the values of $\alpha_\mathrm{M,int}$ and 
$\alpha_\mathrm{M,int}^\mathrm{sat}$.
As $\alpha_\mathrm{M, int}^{\rm sat}$ is proportional 
to $\langle \mu_5\rangle_\mathrm{int}$, for this case the sign of $\langle \mu_5\rangle_\mathrm{int}$ is irrelevant in the expression 
$|\alpha_\mathrm{M} + \eta \langle{\mu_{5}}\rangle|$, 
and we can use the full expressions from 
Eqs.~(\ref{eq_kalpha}) and (\ref{eq_gammaalpha}), which are shown
as dashed-dotted lines in the middle and bottom rows of  
Fig.~\ref{fig_ts_turb_all_meanfield}.
The contribution of $\eta \langle{\mu_{5}}\rangle$ leads to slightly
higher characteristic wave numbers and growth rates, which generally
agree better with the directly measured values of $k_\mathrm{int}$ and $\gamma_\mathrm{int}$.

All of the turbulent runs presented in 
Figs.~\ref{fig_ts_turb_all} and \ref{fig_ts_turb_all_meanfield}
reach saturation eventually, i.e.~the mean magnetic field stops growing.
This can be seen in the time evolution of 
$\langle B\rangle_\mathrm{int}$ in Fig.~\ref{fig_ts_turb_all}  
and in the bottom row of Fig.~\ref{fig_ts_turb_all_meanfield}, 
where $\gamma_\mathrm{int}$ vanishes towards the end of the individual runs.
In case of mean-field dynamos, the maximum magnetic field strength
cannot be estimated by $B_*$ as given by Eq.~(\ref{eq_BCMW}), because
the characteristic timescale is different from that of the small-scale
chiral dynamo.
Instead, we expect 
that the mean-field dynamo instability is saturated by turbulent magnetic diffusion or by nonlinear effects.
In particular, the growth rate of this dynamo vanishes
when $|(\eta+\eta_\mathrm{T}) k_\mathrm{int}| \approx |\alpha_\mathrm{M}
+ \eta \langle{\mu_{5}}\rangle| \approx |\alpha_\mathrm{M}|$.
Indeed, we find that $(\eta+\eta_\mathrm{T}) k_\mathrm{int}$ becomes
comparable to the different estimates of $\alpha_\mathrm{M}$ (see
the top row of  Fig.~\ref{fig_ts_turb_all_meanfield}) at the same
time when $\gamma_\mathrm{int}$ vanishes (see the bottom row of
Fig.~\ref{fig_ts_turb_all_meanfield}).

\begin{figure}[t!]
\centering
  \includegraphics[width=0.45\textwidth]{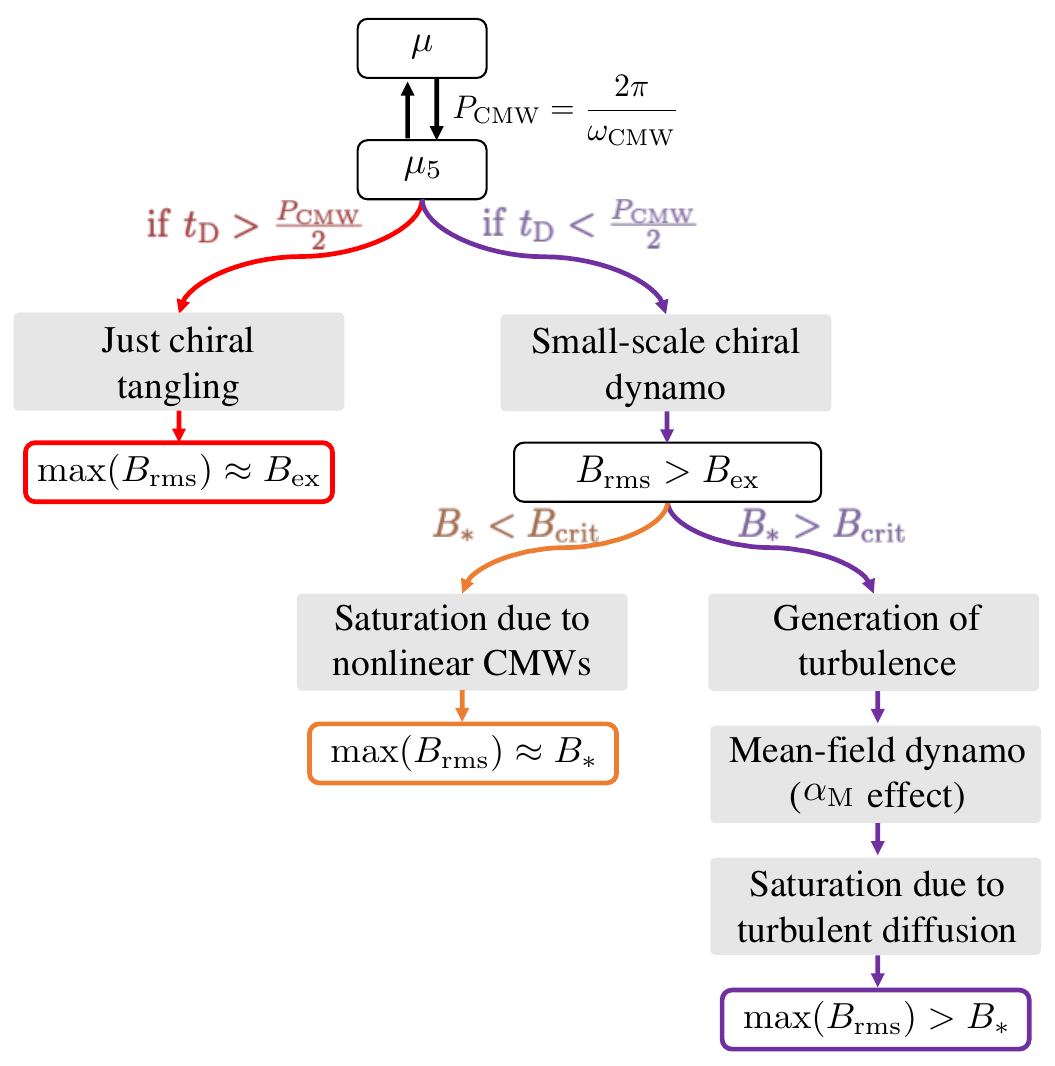} 
  \caption{
Summary of the three different regimes in systems with vanishing initial chiral asymmetry, in which $\mu_5$ is generated through the chiral separation effect. 
The maximum magnetic field strength is the lowest in the regime where just chiral tangling occurs (left side of the sketch). 
When the small-scale chiral dynamo instability is excited,
the maximum field strength is either given by $B_*$ (see Eq.~(\ref{eq_BCMW}); 
middle part of the sketch) or larger if magnetically-driven turbulence is produced
(right side of the sketch).}
  \label{fig_sketch_Bmax}
\end{figure}

\section{Discussion and application to the early Universe}
\label{sec_discussion}

Depending on the parameters and the initial conditions 
of the system, we have identified three different possible evolutionary
branches which are summarized in Fig.~\ref{fig_sketch_Bmax}.
If the timescale of the small-scale chiral dynamo instability
$t_\mathrm{D}$ [see Eq.~(\ref{eq_tD}), and remember that it is based
on the assumption that the total initial $\mu$ can be converted to
$\mu_5$ through the CMW] is smaller than the characteristic period of
the CMW $P_\mathrm{CMW} = 2 \pi/\omega_\mathrm{CMW}$, the magnetic
field fluctuations can only be produced due to chiral tangling (see
Sec.~\ref{sec_chiral_tangling}).
In this case, the maximum magnetic field is limited by the value of the
small imposed magnetic field $B_\mathrm{ex}$.
If $t_\mathrm{D} < P_\mathrm{CMW}$, the small-scale chiral dynamo can
occur and amplify the magnetic fluctuations to values $B_\mathrm{rms}
> B_\mathrm{ex}$.  We have estimated the maximum magnetic field strength
$B_*$ for a given set of initial conditions in Eq.~(\ref{eq_BCMW}), and
find that it depends on the values of the coupling parameters $C_5$
and $C_\mu$, the initial strength and correlation length of the chiral
chemical potential, as well as on the microscopic resistivity $\eta$
and the chiral feedback parameter $\lambda$.
Generally, we expect more efficient magnetic field 
amplification for CMWs with lower frequencies; see 
Fig.~\ref{fig_Bsat_C5Cmu}.
For systems in which $B_* > B_\mathrm{crit}$, the Reynolds number
eventually exceeds unity and the produced turbulence leads to mean-field
effects.
Using DNS, we have shown in
Sec.~\ref{sec_mean_field} that a mean-field dynamo, caused by the magnetic
alpha effect, can amplify the magnetic fluctuations to $B_\mathrm{rms}>B_*$.
We concluded that saturation of the mean-field dynamo is 
caused by an increasing turbulent diffusivity 
$\eta_\mathrm{T}$ in the system. 

The autonomous generation of $\mu_5$ can have consequences for the evolution of a primordial magnetic field until the time when chirality-flipping interactions erase any chiral asymmetry. 
The role of a nonvanishing $\mu_5$ in the early Universe has been 
discussed in many studies, starting with the pioneering work on the
small-scale chiral dynamo 
\cite{JS97}.
Many works on the early Universe 
apply chiral MHD
\cite[e.g.,][]{HiroyukiVachaspatiVilenkin2012,BFR12},
and the highly nonlinear effects caused by a sufficiently amplified magnetic field were characterized \cite{BSRKBFRK17,RogachevskiiEtAl2017, SchoberEtAl2017}.
These studies were based on initial conditions with a nonvanishing $\mu_5$. 
Production of chirality,
however, requires physics beyond the Standard Model and
can, for instance, be realized by the decay of a heavy particle \citep{KamadaEtAl2023}.
In Ref.~\citep{SchoberEtAl2022a}, it was demonstrated that chiral dynamos and the subsequent nonlinear plasma evolution can occur, even if, on average,
there is no chiral asymmetry in the early Universe, but only a spatially fluctuating $\mu_5$. 
In the current study, we report an autonomous generation of these fluctuations of $\mu_5$
in systems with initially vanishing chiral asymmetry if the chemical potential is inhomogeneous and if there is 
a weak uniform magnetic field.

Whether this autonomous generation of $\mu_5$ in the early Universe is sufficient to
lead to a large-scale dynamo instability 
in the primordial magnetic field depends on the 
characteristic parameters of the plasma.
For a large-scale dynamo, the following criteria need to be fulfilled:
(i) An initial weak magnetic field and fluctuations in the chemical potential need to exist to produce $\mu_5$ 
[via the second to last term in Eq.~(\ref{mu5-DNS_CSE})].
(ii) A sufficient separation of scales needs to be established for the small-scale chiral dynamo instability to develop. 
The requirement is that the effective correlation wave number of $\mu_5$ is
$k_\mathrm{\mu_5, eff} \lesssim 5 \mu_{5,\mathrm{max}}$.
Whether this condition is realized or not depends on the initial spectrum of $\mu$, $E_\mu$. 
The amplitude of $E_\mu$ determines the maximum 
possible value of $\mu_{5,\mathrm{max}}$, while the slope of $E_\mu$, 
assuming that it has a power-law shape, determines $k_\mathrm{5, eff}$ 
[remembering that during the linear (in time) phase of $\mu_5$ production, $E_5 \propto k^2 E_\mu$]. 
(iii) The magnetic field produced by the small-scale chiral dynamo 
instability needs to exceed $B_\mathrm{crit}$, 
given by Eq.~(\ref{eq_Bcrit}), 
for the production of turbulence. 
Only if all three conditions are satisfied in the early Universe, 
autonomous generation of $\mu_5$ alone [i.e.~without production of 
$\mu_5$ via physics beyond the Standard Model]
can result in a large-scale dynamo instability.\\ 

\section{Conclusions}
\label{sec_conclusion}

In this paper, we have studied a high-energy plasma with joint action
of the CME and CSE.
We considered a very weak constant initial magnetic field in the 
form of Gaussian fluctuations plus a weak external 
magnetic field $B_\mathrm{ex}$.
The initial chiral chemical potential $\mu_5$ is zero, but there is a
strong initial gradient of the chemical potential fluctuations $\mu$.
Through the CSE, CMWs generate inhomogeneous fluctuations of $\mu_5$.
As there is no (initial) velocity field in the system, the only way
for the magnetic field to get amplified in this scenario is through the
produced chiral asymmetry (i.e., a nonzero $\mu_5$).
The generation of the magnetic field is caused by
the second term on the right-hand side of the induction equation~(\ref{ind-DNS_CSE}). 
However, this term can only lead to a magnetic field instability if the
produced $\mu_5$ becomes large enough.

In this paper, we have identified the parameter space 
in which such an instability can occur. 
Depending on the initial conditions, in particular the
properties of the spatial fluctuations of the 
chemical potential, and the characteristic parameters, 
three different regimes were identified:
(i) a regime in which the magnetic field gets only
amplified through chiral tangling, limiting the maximum 
field strength to that of the imposed field, 
(ii) a regime in which only the small-scale chiral dynamo
occurs,
(iii) a regime in which the small-scale chiral dynamo 
amplifies the magnetic field to high values, such that it drives 
turbulence and a large-scale dynamo instability occurs. 
We found that the large-scale dynamo is best
described by a magnetic alpha effect, and that 
saturation is caused by the buildup of turbulent diffusivity.

With our study, we have shown that chiral dynamo instabilities and even
mean-field dynamos are universal mechanisms for high-energy plasma,
even in the absence of an initial chiral asymmetry.
Our results may have important consequences for the plasma of the early
Universe, protoneutron stars, heavy ion collision experiments,
and the understanding of quantum materials.

\begin{acknowledgements}
This study was initiated several years ago 
through productive discussions with Dmitry Kharzeev. 
We are thankful to the referee for providing constructive comments that improved our paper.
J.S.\ acknowledges the support of the Swiss National Science Foundation under Grant No.\ 185863.
A.B.\ was supported in part through grants from the Swedish Research Council (Vetenskapsr{\aa}det, 2019-04234),
the National Science Foundation under Grant No.\ NSF AST-2307698 and a NASA ATP Award 80NSSC22K0825.
We acknowledge the inspiring atmosphere during the program on ``Turbulence in Astrophysical Environments'' at the Kavli Institute for Theoretical
Physics in Santa Barbara, supported by the National Science Foundation under Grant No.\ NSF PHY-2309135.
I.R.\ would like to thank the Isaac Newton Institute for Mathematical Sciences, Cambridge University, for support and hospitality during the program
``Anti-diffusive dynamics: from sub-cellular to astrophysical scales'', where the final version of the paper was completed.
\end{acknowledgements}

\bigskip

\appendix 

\section{Justification of Eqs.~(\ref{ind-DNS_CSE})--(\ref{mu5-DNS_CSE})}
\label{sec_justification}

In Sec.~\ref{ChiralMHD}, we stated the governing equations used in this paper.
Here we provide more background regarding their derivation.
The continuity equations for the number densities $n_5=n_{\rm L} - n_{\rm R}$ and $n=n_{\rm L} + n_{\rm R}$
(which are proportional to the chiral chemical potential $\mu_5^{\rm phys}=\mu^{\rm phys}_{\rm L} - \mu^{\rm phys}_{\rm R}$
and the chemical potential $\mu^{\rm phys}=\mu^{\rm phys}_{\rm L} + \mu^{\rm phys}_{\rm R}$, respectively) are given by
\begin{eqnarray}
\frac{\partial n_5}{\partial t} + {\bm \nabla} \cdot \left[n_5 \UU + \frac{e}{2 \pi^2 \hbar^2 \, c} \,  \mu^{\rm phys} \BB\right]=
\frac{e^2}{2\pi \hbar^2 \, c} \, \EE\cdot \BB ,
\nonumber\\
  \label{eq-n5}
\end{eqnarray}
\begin{eqnarray}
\frac{\partial n}{\partial t} + {\bm \nabla} \cdot \left[n \UU + \frac{e}{2 \pi^2 \hbar^2 \, c} \,  \mu_5^{\rm phys} \BB + \frac{\sigma}{e} \EE \right]=
0 ,
  \label{eq-n}
\end{eqnarray}
where $n_{\rm L}$ and $n_{\rm R}$ are the number densities of the left- and right-handed electrically
charged fermions, respectively, 
$\mu^{\rm phys}_{\rm L}$  and $\mu^{\rm phys}_{\rm R}$ 
are the chemical potentials of the left- and right-handed electrically
charged fermions,
$e$ is the electric charge, $\hbar$ is Planck's constant, $c$ is the speed of light, 
$\EE$ is the electric field, $\BB$ is the magnetic field, $\UU $ is the plasma velocity,
and $\sigma$ is the electric conductivity of plasma.
The second term in the squared brackets of Eq.~(\ref{eq-n5}) describes the chiral separation effect \citep{SonZhitnitsky2004},
while the second and third terms in the squared brackets of Eq.~(\ref{eq-n}) determine
the chiral magnetic effect \citep{Vilenkin:80a} and the electric charge screening effect \citep{RGS19}, respectively.

Equations~(\ref{eq-n5})--(\ref{eq-n}) are written in the Heaviside-Lorentz system of units where $c=1$.
In the present paper we use Gaussian units 
(in accordance with most of the literature in plasma physics and astrophysics),
so that the coefficient $e^2/(2\pi^2 \hbar^2 c)$ should be replaced by
$2e^2/(\pi \hbar^2 c)$, where $\alphaem \equiv e^2/(\hbar c) \approx 1 /137$ is the fine-structure constant.
Now we define the normalized chiral chemical potential $\mu_5$ and chemical potential $\mu$ as
$\mu_5  = 4 \mu_5^{\rm phys} \alphaem / (\hbar c)$ and $\mu  = 4 \mu^{\rm phys} \alphaem / (\hbar c)$, so that
our new variables $\mu_5$ and $\mu$ have the dimension of inverse length.

Since the main focus of the paper is the effect of the chiral asymmetry production by inhomogeneous fluctuations of chemical potential and since
the chiral dynamo effect and the production of turbulence studied in the present paper develop on a timescale which is less than a half period of the CMWs, we neglect 
the electric charge screening which causes a damping of the CMWs \citep{RGS19}.
Thus, Eqs.~(\ref{eq-n5})--(\ref{eq-n}) yield Eqs.~(\ref{mu-DNS_CSE})--(\ref{mu5-DNS_CSE}),
where for numerical stability we also added hyperdiffusion terms
with the diffusion coefficients $\mathcal{D}_5$ and $\mathcal{D}_\mu$
\citep{SchoberEtAl2022b}.

We consider a system which consists of a nonrelativistic plasma whose
electric properties are described by the Ohmic current and the
electric charge density.
The nonrelativistic dynamics of the plasma is governed
by the Maxwell equations and the Navier-Stokes equation
relating the fluid velocity, $|\UU| \ll c$, to the magnetic field, $\BB$.
The nonrelativistic plasma interacts with highly relativistic electrically charged fermions.
The electric current, $\propto \mu_5 \BB$, caused by the relativistic plasma component, is an additional source for the magnetic field in the
Maxwell equations (see the detailed discussions related to different
plasma models in Ref.~\cite{RogachevskiiEtAl2017}).
The electric field for very small magnetic diffusion $\eta=c^2/4 \pi \sigma$ (typical for astrophysical systems with large magnetic Reynolds numbers)
is given by \citep{RogachevskiiEtAl2017}:
\begin{eqnarray}
\EE &=& - \frac{1}{c} \, \biggl[\UU\times \BB + \eta \left(\mu_5 \BB - {\bm \nabla}\times \BB\right)\biggr]
+ O(\eta^2) .
  \label{eqE}
\end{eqnarray}
The magnetic field $\BB$ is normalized such that the magnetic energy density is $\BB^2/2$ without the $4\pi$ factor.
MHD is formulated as the evolution of the magnetic and velocity fields,
neglecting the Faraday displacement current in the Maxwell equation for
${\bm \nabla}\times \BB$.
Substituting the electric field $\EE$ given by Eq.~(\ref{eqE}) 
in the Maxwell equation for $\partial \BB/ \partial t$, we obtain the induction equation~(\ref{ind-DNS_CSE}) for the chiral MHD.

In the nonlinear stage of the chiral dynamo instability,
the velocity fluctuations are produced by the Lorentz force in the Navier-Stokes equation.
The plasma motions with the bulk velocity $\UU$ are described by the Navier-Stokes equation~(\ref{UU-DNS_CSE}) and continuity equation~(\ref{rho-DNS_CSE}) which coincide with corresponding equations of the classical MHD \citep{RogachevskiiEtAl2017}.

\section{Numerical constraints for simulations with CMWs}
\label{sec_DNScriteria}

In the simulations presented in this study, two crucial criteria need to be satisfied.
As in any simulation of chiral MHD, the resolution
needs to be high enough to resolve the small-scale chiral instability.
The instability is attained on the wave number $k_5$ given in Eq.~(\ref{eq_k5}).
With the minimum wave number in the numerical domain with
resolution $N$ being $2\pi N/ L$,
the criterion for chiral MHD simulations is 
\begin{eqnarray}
   \frac{2\pi N}{L} \gtrsim \frac{\mu_5}{2}.
\label{eq_Ncrit}
\end{eqnarray}
If $\mu_5$ is produced from CMWs, the approximate maximum
value of $\mu_5$ is the initial value of the chemical potential, $\mu(t_0)$,
and therefore the criterion in Eq.~(\ref{eq_Ncrit}) becomes
\begin{eqnarray}
   \frac{2\pi N}{L} \gtrsim \frac{\mu(t_0)}{2}.
\label{eq_Ncrit2}
\end{eqnarray}

Another constraint on the parameter space that is 
accessible with DNS is related to the time step.
As discussed in \citep{SBR20}, the time step contribution from 
the terms including $\mu_5$ and $\mu$ is
\begin{eqnarray}
   \delta t_\mathrm{chiral} = c_\mathrm{\delta t,chiral} ~
     \mathrm{min} (\delta t_{\lambda_5}, \delta t_{D_5}, 
     \delta t_{\mathrm{CMW}}, \delta t_{D_\mu}, \delta t_{v_\mu}) \nonumber \\
\end{eqnarray}
with
\begin{align}
   \delta t_{v_\mu} = \,&\frac{\delta x}{\eta \mu_5},\,&
   \delta t_{\mathrm{CMW}} = \,&\frac{\delta x}{|\BB|\sqrt{C_5 C_\mu}},\,&
\delta t_{\lambda_5} = \,&\frac{1}{\lambda \eta \BB^2}, \nonumber  \\
   \delta t_{D_5} = \,&\frac{\delta x^4}{\mathcal{D}_5}   ,\,&
   \delta t_{D_\mu} =\,& \frac{\delta x^4}{\mathcal{D}_\mu},\,&
\end{align}
and with the scaling parameter 
$c_\mathrm{\delta t,chiral}$.
For CMWs with large frequencies, the contribution from
$\delta t_{\mathrm{CMW}}$ becomes the most relevant one. 
With $B$ increasing through the chiral dynamo instability, 
the CMW frequency increases, in other words, the 
characteristic velocity of the CMWs
\begin{eqnarray}
    v_\mathrm{CMW} \approx |\BB|\sqrt{C_5 C_\mu}
\label{eq_vCMW}
\end{eqnarray}
becomes larger.
If $v_\mathrm{CMW}$ becomes larger than the sound speed
$c_\mathrm{s}=1$, shocks develop and the numerical solution
becomes unstable. 
Therefore, the parameters should be chosen such that 
the maximum magnetic field strength $B_*$ generated self-consistently 
through CMWs [see Eq.~(\ref{eq_vCMW})],
is less than $(C_5C_\mu)^{-1/2}$.

\end{document}